\pdfoutput=1

\documentclass{article}

\usepackage{arxiv}

\usepackage[utf8]{inputenc} % allow utf-8 input
\usepackage[T1]{fontenc}    % use 8-bit T1 fonts
\usepackage{lmodern}        % https://github.com/rstudio/rticles/issues/343
\usepackage{hyperref}       % hyperlinks
\usepackage{url}            % simple URL typesetting
\usepackage{booktabs}       % professional-quality tables
\usepackage{amsfonts}       % blackboard math symbols
\usepackage{nicefrac}       % compact symbols for 1/2, etc.
\usepackage{microtype}      % microtypography
\usepackage{graphicx}

\title{An open-source integrated framework for the automation of
citation collection and screening in systematic reviews.}

\author{
    Angelo D'Ambrosio
   \\
    Institute for Infection Prevention and Hospital Hygiene \\
    Freiburg University Hospital \\
  Freiburg, Germany \\
  \texttt{\href{mailto:angelo.d.ambrosio@uniklinik-freiburg.de}{\nolinkurl{angelo.d.ambrosio@uniklinik-freiburg.de}};
\href{mailto:a.dambrosioMD@gmail.com}{\nolinkurl{a.dambrosioMD@gmail.com}}} \\
   \And
    Hajo Grundmann
   \\
    Institute for Infection Prevention and Hospital Hygiene \\
    Freiburg University Hospital \\
  Freiburg, Germany \\
  \texttt{\href{mailto:hajo.grundmann@uniklinik-freiburg.de}{\nolinkurl{hajo.grundmann@uniklinik-freiburg.de}}} \\
   \And
    Tjibbe Donker
   \\
    Institute for Infection Prevention and Hospital Hygiene \\
    Freiburg University Hospital \\
  Freiburg, Germany \\
  \texttt{\href{mailto:tjibbe.donker@uniklinik-freiburg.de}{\nolinkurl{tjibbe.donker@uniklinik-freiburg.de}}} \\
  }

\providecommand{\tightlist}{%
  \setlength{\itemsep}{0pt}\setlength{\parskip}{0pt}}

\newlength{\cslhangindent}
\newlength{\csllabelwidth}
\newlength{\cslentryspacingunit} % times entry-spacing
  {}%
  {\par}
\newenvironment{CSLReferences}[2] % #1 hanging-ident, #2 entry spacing
 {% don't indent paragraphs
  \setlength{\parindent}{0pt}
  \ifodd #1
  \let\oldpar\par
  \def\par{\hangindent=\cslhangindent\oldpar}
  \fi
  \setlength{\parskip}{#2\cslentryspacingunit}
 }%
 {}
\usepackage{calc}

\usepackage{caption}
\usepackage{placeins}
\usepackage{amsmath}
\usepackage{booktabs}
\usepackage{longtable}
\usepackage{array}
\usepackage{multirow}
\usepackage{wrapfig}
\usepackage{float}
\usepackage{colortbl}
\usepackage{pdflscape}
\usepackage{tabu}
\usepackage{threeparttable}
\usepackage{threeparttablex}
\usepackage[normalem]{ulem}
\usepackage{makecell}
\usepackage{xcolor}
\begin{document}
\maketitle

\begin{abstract}
The exponential growth of scientific production makes secondary
literature abridgements increasingly demanding. We introduce a new
open-source framework for systematic reviews that significantly reduces
time and workload for collecting and screening scientific literature.\\
The framework provides three main tools: 1) an automatic citation search
engine and manager that collects records from multiple online sources
with a unified query syntax, 2) a Bayesian, active machine learning,
citation screening tool based on iterative human-machine interaction to
increase predictive accuracy and, 3) a semi-automatic, data-driven query
generator to create new search queries from existing citation data
sets.\\
To evaluate the automatic screener's performance, we estimated the
median posterior sensitivity and efficiency {[}90\% Credible
Intervals{]} using Bayesian simulation to predict the distribution of
undetected potentially relevant records.\\
Tested on an example topic, the framework collected 17,755 unique
records through the citation manager; 766 records required human
evaluation while the rest were excluded by the automatic classifier; the
theoretical efficiency was 95.6\% {[}95.3\%, 95.7\%{]} with a
sensitivity of 100\% {[}93.5\%, 100\%{]}.\\
A new search query was generated from the labelled dataset, and 82,579
additional records were collected; only 567 records required human
review after automatic screening, and six additional positive matches
were found. The overall expected sensitivity decreased to 97.3\%
{[}73.8\%, 100\%{]} while the efficiency increased to 98.6\% {[}98.2\%,
98.7\%{]}.\\
The framework can significantly reduce the workload required to conduct
large literature reviews by simplifying citation collection and
screening while demonstrating exceptional sensitivity. Such a tool can
improve the standardization and repeatability of systematic reviews.
\end{abstract}


\hypertarget{introduction}{%
\section{Introduction}\label{introduction}}

Scientific production has experienced continuous exponential growth in
the last decades (Larsen and Von Ins 2010; Bornmann and Mutz 2015). This
is especially true for biomedical research, a trend further accelerated
by the COVID-19 pandemic, thanks to faster article processing time by
publishers and the greater use of preprint databases (Aviv-Reuven and
Rosenfeld 2021; Horbach 2020; Hoy 2020). Consequently, it has become
harder for researchers and practitioners to stay up to date on the
latest findings in their field. Secondary research is of paramount
importance in this scenario in that it provides valuable summaries of
the latest research results; however, it is becoming ever more
challenging in terms of time and human resources required (Allen and
Olkin 1999; Borah et al. 2017; A. M. Cohen et al. 2010; Bastian,
Glasziou, and Chalmers 2010).\\
The article collection and screening phases of a systematic review are
particularly demanding (Babar and Zhang 2009). First, relevant published
research must be collected from scientific databases using appropriately
built search queries (retrieval phase); secondly, the scientific
citations collected must be screened, selecting only those that are
relevant to the topic (appraisal phase) (Bannach-Brown et al. 2019;
Tsafnat et al. 2014; Higgins et al. 2019).\\
Search queries construction is a complex task (Lefebvre et al. 2011;
Hammerstrøm et al. 2010), requiring both expertise in the scientific
field of interest and some knowledge of the database query languages.
The goal is to obtain a set of results that contains all relevant
articles (high sensitivity) while keeping the total number of records
low (high specificity), possibly focusing on the first at the expense of
the second (Hammerstrøm et al. 2010).\\
If an integrated search tool is not used, manual work is required to
download, store and organise the publication data; this approach is
complicated by limits to the number of records that can be downloaded at
any one time and the need to harmonise different formats and resolve
record duplication (Marshall and Wallace 2019).\\
The citation screening phase is usually the more resource-demanding task
in a systematic review: even with appropriately built search queries,
the results may easily range in the tens of thousands, of which just a
small fraction are actually relevant (Lefebvre et al. 2011). It has been
estimated that labelling 10,000 publications can take up to 40 weeks of
work and that the average clinical systematic review takes 63 weeks to
complete (Bannach-Brown et al. 2019; Borah et al. 2017; Allen and Olkin
1999). A consequence of this is that systematic reviews are often
already out-of-date by the time they are published (E. M. Beller et al.
2013).\\
The field of Data Science applied to evidence synthesis and acquisition
has greatly maturated in the last years (Marshall and Wallace 2019; E.
Beller et al. 2018; Tsafnat et al. 2014). By applying natural language
processing (NLP), it is possible to transform free text into
quantitative features, with various levels of abstraction and
generalisation (Ananiadou and McNaught 2006; K. B. Cohen and Hunter
2008); using machine learning, such text-derived data can be used to map
and reproduce human judgment, automating the screening of citations
(Ikonomakis, Kotsiantis, and Tampakas 2005).\\
Automation of systematic reviews has made significant improvements in
the last years (Ananiadou et al. 2009; O'Mara-Eves et al. 2015; Tsafnat
et al. 2013; Jonnalagadda, Goyal, and Huffman 2015), and it is possible
foreseeable that it will become the standard approach in the field (E.
Beller et al. 2018), with many solutions already being implemented into
commercial or free-to-use tools (see Marshall and Wallace 2019, table
1).\\
This manuscript introduces an open-source, production-ready framework
that further contributes to the state-of-the-art in systematic review
automation (SRA) and helpers (SRH) tools. We improve the ``retrieval
phase'' by providing a unified framework for the automated collection
and management of scientific literature from multiple online sources.
For the citation screening (appraisal) phase, we built an active machine
learning-based protocol (Settles 2009; Miwa et al. 2014), which utilises
a Bayesian framework to efficiently identify potentially relevant
documents that require human review while automatically screening-out
the vast majority of clearly non-relevant ones; the algorithm then
requires human review to increase classification accuracy iteratively.
Finally, we included a tool to generate new search queries based on an
already categorised citation data set, to identify relevant research
that manually-made queries may have possibly missed.\\
We tested the framework in the retrieval and appraisal phases of an
example topic of interest to our group: the evaluation of the
mathematical modelling of patient referral networks among hospitals and
their impact on the diffusion of healthcare-associated pathogenic
microorganisms; the protocol is published in (Sadaghiani et al. 2020).\\
In the Methods, we give an overview of the framework, in the Result, we
show the outputs and performance of the framework applied to the example
topic, and in the Discussion, we explain the methodological rationale
for the different components and features of the framework.\\

\hypertarget{methods}{%
\section{Methods}\label{methods}}

\hypertarget{general-description}{%
\subsection{General description}\label{general-description}}

We built an R (R Core Team 2020) based framework to simplify two aspects
of systematic literature review: record acquisition and classification.
The code used to generate the results is available at
\url{https://github.com/AD-Papers-Material/BART_SystReviewClassifier},
while an updated and ready to use version of the framework is
distributed as an R package at
\url{https://github.com/bakaburg1/BaySREn}. The framework includes
several modules that communicate through intermediate outputs stored in
standard formats, making it possible for users to extend the framework
or easily integrate it with other tools in their pipeline. See
Supplemental Material S1 for an in-depth description of the framework
and how to use it.\\
The tasks carried out by the framework are grouped into ``sessions'',
which comprise obtaining scientific citation data (records) using a
search query and then labelling them as relevant (``positive'' in the
rest of the text) or not (``negative'') for the topic of interest with
the help of a machine learning engine (Fig. 1). The initial search query
should be built using domain knowledge, trying to achieve a high
relevant/non-relevant record ratio.\\
The framework can then generate a new data-driven query from this
labelled set to perform a new session to find records possibly missed by
the first query.\\

\begin{figure}
\includegraphics[width=1\linewidth]{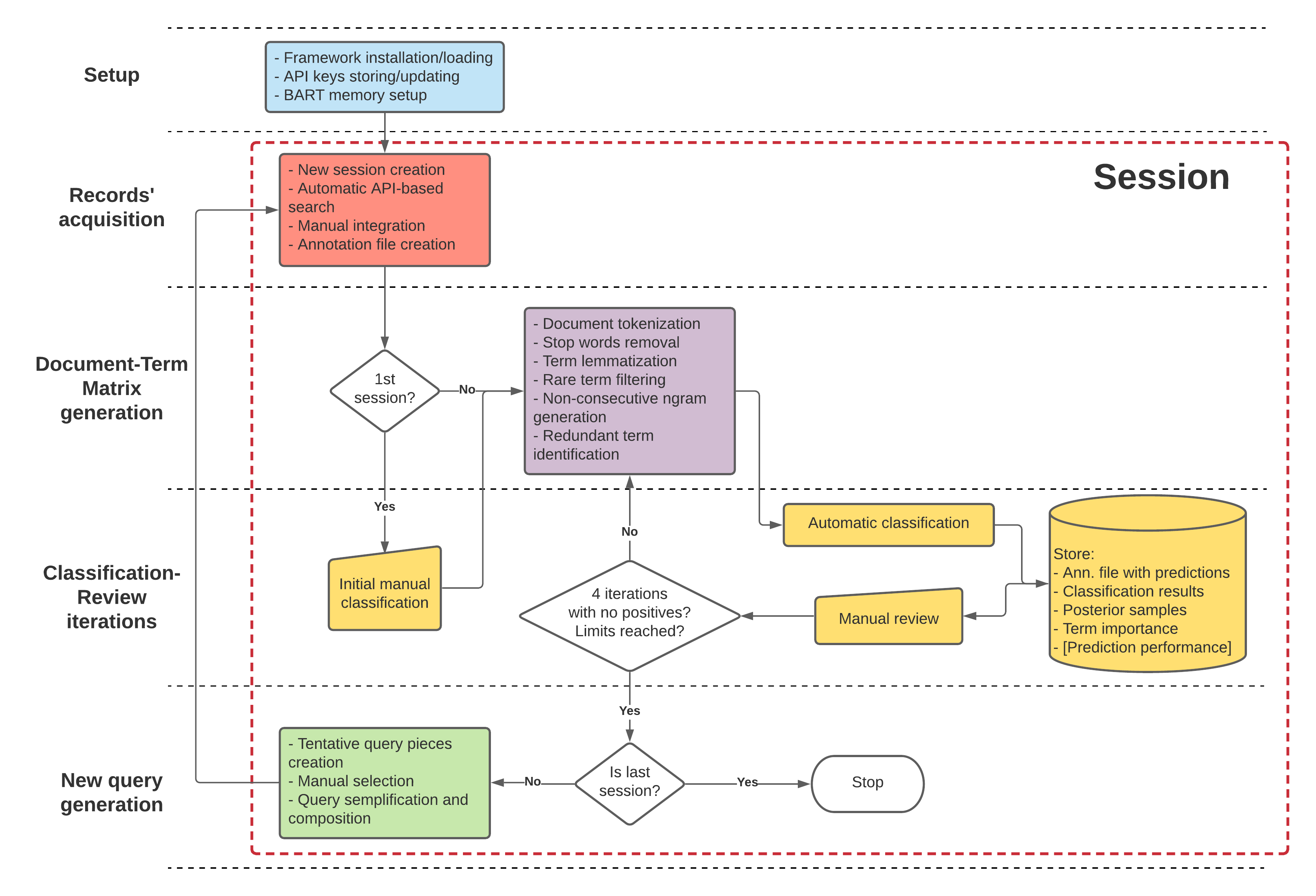} \caption{Figure 1. Framework's visual representation.}\label{fig:method_diagram}
\end{figure}

\hypertarget{record-acquisition-and-initial-labelling}{%
\subsection{Record acquisition and initial
labelling}\label{record-acquisition-and-initial-labelling}}

We built a set of tools to allow users to automatically search and
download citation data from three major scientific databases
(``sources''): Pubmed (\url{https://pubmed.ncbi.nlm.nih.gov/}), Web Of
Science (WOS, \url{https://apps.webofknowledge.com/}) and the Institute
of Electrical and Electronics Engineers (IEEE,
\url{https://ieeexplore.ieee.org/Xplore/home.jsp}). The framework
handles authorisation management for non-open databases like WOS and
IEEE. It is also possible to import previously downloaded records in the
framework; this is particularly useful for acquiring records from SCOPUS
(\url{https://www.scopus.com/}) and EMBASE databases
(\url{https://www.embase.com/}), for which a comprehensive API interface
was not easy to build. An extra manual search was also necessary for
Pubmed since the API and the web interface have different rule expansion
algorithms and return slightly different results ({`NCBI Insights :
Updated Pubmed e-Utilities Coming in April 2022!'}, n.d.). A short guide
on how to set up the framework for each database supported is available
in Supplemental Material S3.\\
The collected records are merged into a single database, resolving
duplicates and different formatting between sources. The records are
ordered according to the frequency of the positive query terms (e.g.,
not preceded by a \emph{NOT} modifier) in the title and abstract
(``simple query ordering'').\\
The researcher is then asked to label a subset of records to create the
``initial training set'' needed to start the automatic classification.
We recommend manually labelling the first 250 records (see
``hyperparameter optimisation'' later). Simple query ordering increases
the positivity rate in the initial training set (Wallace, Small, et al.
2010), leading to higher sensitivity during automatic classification
(Chawla, Japkowicz, and Kotcz 2004).

\hypertarget{text-feature-extraction}{%
\subsection{Text feature extraction}\label{text-feature-extraction}}

The framework models the relevance of a record based on the following
fields in the citation data: title, abstract, authors, keywords, MESH
terms (Lipscomb 2000). A range of Natural Language Processing (NLP)
techniques (Baeza-Yates, Ribeiro-Neto, et al. 1999; Marshall and Wallace
2019; Ananiadou and McNaught 2006) are employed to convert the textual
information in these fields into features for machine learning through a
bag-of-words approach (Marshall and Wallace 2019). Processing of free
text fields (title, abstract) includes: tokenisation (i.e., extracting
the terms), removal of common stopwords (i.e., sentence components
having no semantic value), part-of-speech filtering (only nouns,
adjectives, verbs and untagged terms are retained), and lemmatisation of
terms (i.e., reduction to their base grammatical form). Text processing
for authors, keywords and MESH terms identifies logical units (e.g.,
authors' full names, composite keywords) and extracts them.\\
Terms appearing in less than 5\% of the labelled documents are removed
from negative records. All terms in the positive set are kept to
increase sensitivity at the cost of specificity.\\
Some terms tend to co-appear in records (non-consecutive ngrams,
nc-ngrams), often carrying a particular meaning when they do co-occur.
To detect nc-ngrams, we generated a word network representation
(Francois Rousseau 2015) with edges occurring between terms with a
cosine similarity in terms of document co-occurrence \textgreater{} 0.5.
We extracted the maximal cliques in the network (Eppstein, Löffler, and
Strash 2010) representing highly correlated groups of terms; these
groups are added to the dataset as individual features. Only nc-ngrams
comprising a maximum of ten terms are kept.\\
A second network is built using a co-occurrence threshold of 0.9. In
this case, the cliques represent terms that always appear together and
can therefore be considered redundant (i.e., they do not need to be
considered separately). These terms are merged to increase computational
efficiency and reduce overfitting.\\
The output is a Document-Term Matrix (DTM), with \(N_d\) rows
representing the records (\(D_i\)), \(N_t\) terms column for the
\(t_{field}\) terms (divided by record field) and \({0,1}\) values
whether \(t_{field} \in D_i\). We also enriched the DTM with features
referencing the number of terms in each field to help the model scale
term importance based on the field length.

\hypertarget{label-prediction}{%
\subsection{Label prediction}\label{label-prediction}}

We used a Bayesian Additive Regression Trees (BART) machine learning
``classification model'' (Chipman et al. 2010) (in the implementation of
Kapelner and Bleich 2013) to predict the probability of a record being
relevant, given the information coded into the enriched DTM and the
initial training set. We set up the BART model to use 2,000 MCMC
iterations (after 250 burn-in iterations) and 50 trees; we used a \(k\)
value of 2 to regularise extreme prediction and let the model use
missing fields in the DTM as features (Kapelner and Bleich 2015).
Positive records are oversampled ten times to increase sensitivity
(Batista, Prati, and Monard 2004).\\
The output is a posterior predictive distribution (PPD) for each record
describing the probability of it being relevant (i.e., a positive
match). An ensemble of ten models was fitted to improve prediction
stability by averaging the PPD between models (Zhou 2021; Dietterich
2000).\\

To assign the labels, we employed an ``active learning'' approach
(Settles 2009; Miwa et al. 2014), where a human reviews a specific
subset of predictions made by the machine, which is then retrained on
the manually reviewed dataset. This process is carried out iteratively
to reduce prediction uncertainty.\\
Label assignment is done through identification of an ``uncertainty
zone'', the construction of which is possible thanks to the Bayesian
nature of BART, which provides full PPDs instead of point-wise
predictions for each record.\\
To describe the process formally, we define:

\[\pi_i = \frac{1}{M}\sum_{j=1}^M Pr(L_i = \text{1}|DTM,m_j)\]

as the PPD of a record \(D_i\) being relevant (i.e, having a positive
label, \(L_i = 1\)), averaging the PPDs of the ensemble of \(M=10\)
models \(m\), and:

\[
\begin{aligned}
\pi_{i,l} = \{\pi_i : Pr(\pi_i) = 1\%\}\\
\pi_{i,u} = \{\pi_i : Pr(\pi_i) = 99\%\}
\end{aligned}
\]

respectively as the lower and upper boundaries of the 98\% quantile
interval of \(\pi_i\) (98\% Predictive Interval, 98\% PrI).\\
Then we identify the ``uncertainty zone'' as:

\[U_\pi=[\max\vec{\pi}_{u}^-, \min\vec{\pi}_{l}^+]\]

with \(\vec{\pi}_{u}^-\) being the vector of \(\pi_{i,u}\) with a
negative label and \(\vec{\pi}_{l}^+\) the vector of \(\pi_{i,l}\) with
a positive label. That is, \(U_\pi\) defines a range of values between
the smallest \(\pi_{i,l}\) in the set of already labelled positive
records \(L_p\) and the largest \(pi_{i,u}\) related to the negative
ones \(L_n\), noting that the two limits can appear in any order.\\
Consequently, a record \(D_i\) will be labelled as positive if:

\[\pi_{i,l} > \max_{\pi \in U_\pi} \pi\]

that is, the record lower 98\% PrI boundary should be higher than every
value in the uncertainty zone. In other words, for a record to be
labelled positive, its PPD should be within the range of the mixture of
PPD of the records previously labelled positive and should not cross the
distributions of the negative records.\\
Conversely, a record is labelled as negative if:

\[\pi_{i,u} < \min_{\pi \in U_\pi} \pi\]

All other records are labelled as ``uncertain''.

Manual review is then necessary for: 1) uncertain records, 2) positive
records (to avoid false positives), and 3) records whose predicted label
differs from the existing manual one. The last case helps identify human
errors or inconsistent labelling criteria.

The automatic classification and manual review steps alternate in a loop
(CR iterations) until no new positive matches are found in four
consecutive iterations.

\hypertarget{relevant-term-extraction}{%
\subsection{Relevant term extraction}\label{relevant-term-extraction}}

As a measure of feature importance, we computed the ``inclusion rate'',
that is, the proportion of times a term is used in a posterior tree over
the sum of total inclusions of all variables (Kapelner and Bleich 2013).
We extracted the terms, the portion of the citation data in which they
were used, the average ``inclusion rate'' among the ensemble models
(over 10,000 inclusions) and its ratio over the standard deviation of
this inclusion (inclusion stability, IS). For each term, we ran a
Poisson regression to get the linear association with a positive label
and reported it as Relative Risk (RR) with the number of standard errors
as significance index (Statistic); the comparison between the inclusion
rate in the BART models and the linear association allows to spot
relevant non-linear effects (i.e., the feature is relevant only in
association with others). In the Results, we only listed the first 15
terms with IS \textgreater{} 1.5 (in order of inclusion rate), while the
first fifty terms, regardless of inclusion stability, are listed in
Supplemental Material S2.

\hypertarget{new-search-query-generation}{%
\subsection{New search query
generation}\label{new-search-query-generation}}

We developed an algorithm that generates a new search query to find
further relevant publications missed in the initial search, possibly at
a reasonable cost to specificity (i.e., a higher number of negative
results).\\
The algorithm is composed of the following steps:

\begin{itemize}
\tightlist
\item
  A partition tree (Therneau and Atkinson 2019) is built between the DTM
  and 800 samples from the PPD; if a term is present multiple times in
  the DTM (e.g., both in the title and abstract), it is counted just
  once, and field term count features are removed. This step generates a
  list of rules composed by \emph{AND}/\emph{NOT} ``conditions'' made of
  terms/authors/keywords/MESH tokens, which together identify a group of
  records.
\item
  For each rule, negative conditions (i.e., \emph{NOT} statements) are
  added iteratively, starting from the most specific one, until no
  conditions are found that would not also remove positive records.
\item
  The extended set of rules is sorted by positive-negative record
  difference in descending order. The cumulative number of unique
  positive records is computed and used to group the rules. Rules inside
  each group are ordered by specificity.
\item
  The researcher is then asked to review the rule groups and select one
  or more rules from each group or edit overly specific rules (e.g.,
  citing a non-relevant concept casually associated with a paper, like a
  numeric value or indicator). It is possible to exclude a group of
  rules altogether, especially those with the poorest
  sensitivity/specificity ratio.
\item
  The selected rules are joined together by \emph{OR} statements,
  defining a subset of records with a sensibly higher proportion of
  positive records than the original set
\item
  Redundant (i.e., rules whose positive records are already included in
  more specific ones) and non-relevant rules (i.e., conditions that when
  removed do not impact sensitivity and specificity) are removed.
\item
  Finally, the rules are re-elaborated in a query that can be used to
  perform a new citation search.
\end{itemize}

Because the algorithm is data-driven, it creates queries that
effectively select positive records from the input dataset but may be
not specific enough when applied to actual research databases. Therefore
we added an extra subquery in \_AND\_that specifies the general topics
of our search and narrows the search domain.\\
The new query was used to initiate a second search session.

\hypertarget{performance-evaluation}{%
\subsection{Performance evaluation}\label{performance-evaluation}}

We trained a simple Bayesian logistic regression (surrogate model) on
the reviewed records to evaluate the consistency of the classification
model (see Discussion for the theoretical justification). The surrogate
model uses as predictor the lower boundary of the 98\% PrI of the PPD of
the records with weakly regularising, robust priors for the intercept
(Student T with \(\nu=3,\mu=0,\sigma=2.5\)) and the linear coefficient
(Student T with \(\nu=3,\mu=0,\sigma=1.5\)).\\
The quality of the model was evaluated through the Bayesian \(R^2\)
(Gelman et al. 2019), of which we reported the posterior median and 90\%
Credible Interval {[}90\% CrI{]}. The \(R^2\) also provides an
evaluation of the consistency of the original classification model.
Given that this model is conditional only on the BART predictions and
not on the DTM, it is characterised by more uncertainty, providing
plausible worst-case scenarios.\\
The surrogate model is then used to generate the predictive cumulative
distribution of the number of total positive records in the whole
dataset. This distribution allows estimating the expected total
posterior ``Sensitivity'' and ``Efficiency'' of the classification model
in the whole (unreviewed) dataset. Efficiency is summarised by the
``Work saved over random'' (WSoR) statistic: one minus the ratio between
the number of records manually reviewed and those that would be required
to find the same number of positives if classification were performed
choosing records randomly; this last quantity is estimated through a
negative hypergeometric distribution (Chae 1993) over the predicted
number of positive records.\\
For the number of predicted positive records, sensitivity and
efficiency, we reported the ``truncated 90\% PrI'' {[}trunc. 90\%
PrI{]}, i.e., the uncertainty interval bounded by the number of observed
total positive records (i.e., there cannot be fewer predicted positive
records than observed).

\hypertarget{hyperparameter-evaluation}{%
\subsection{Hyperparameter evaluation}\label{hyperparameter-evaluation}}

Our classification algorithm has a limited number of hyperparameters:

\begin{itemize}
\tightlist
\item
  Size of the initial training set: 50, 100, 250, 500 records;
\item
  Number of models in the ensemble: 1, 5, 10, 20, 40, 60 repetitions;
\item
  Oversampling rate of positive records: (1x (i.e., no oversampling),
  10x, 20x;
\item
  PrI quantiles for building the uncertainty zone: 80\%, 90\%, 98\%;
\item
  Source of randomness between models in the ensemble: MCMC sampling
  only (Robert and Casella 2004), MCMC plus data bootstrapping (Breiman
  1996) of the training set.
\end{itemize}

To evaluate the hyperparameter effect of performance, we set up a ``grid
search'' (Claesen and De Moor 2015; L. Yang and Shami 2020) on a
prelabelled ``validation set'' derived from the first 1,200 records of
the first session dataset. Each hyperparameter combination was tested
until four CR iterations were completed with no positive records or
until the whole dataset was labelled.\\
For each combination, a performance score was computed as the product of
``Efficiency'' (1 minus the ratio of records that required reviewing
over the total number of records) and ``Sensitivity'' (number of
positive records found over the total number of positive records). We
then used a partition tree (Therneau and Atkinson 2019) to identify
homogeneous ``performance clusters'' of scores given hyperparameter
values. For the rest of the study, we used the best hyperparameter set
in terms of sensitivity followed by efficiency from the cluster with the
highest average score.\\

\hypertarget{results}{%
\section{Results}\label{results}}

\hypertarget{first-session}{%
\subsection{First session}\label{first-session}}

The initial search query for the example topic was:\\

\emph{((model OR models OR modeling OR network OR networks) AND
(dissemination OR transmission OR spread OR diffusion) AND (nosocomial
OR hospital OR ``long-term-care'' OR ``long term care'' OR ``longterm
care'' OR ``long-term care'' OR ``healthcare associated'') AND
(infection OR resistance OR resistant))}

selecting only results between 2010 and 2020 (included). Results were
collected from Pubmed, WOS, IEEE, EMBASE and SCOPUS, using the framework
tools as described in the Methods and Supplemental Material S1.

The first search session returned a total of 27,600 records,
specifically 12,719 (71.6\% of the total) records from the EMBASE
database, followed by 9,546 (53.8\%) from Pubmed, 3,175 (17.9\%) from
SCOPUS, 2,100 (11.8\%) from WOS, and 60 (0.34\%) from IEEE (Table 1).
There were various degrees of overlapping between sources, with 38.4\%
of records being present in more than one database, and EMBASE and IEEE
being the databases with the higher uniqueness ratios. The final data
set was composed of 17,755 unique records.\\
The first 250 records (based on ``simple query ordering'') were
categorised manually. Of these 43 (17.2\%) were labeled as positive, and
207 (82.8\%) as negative.

\begin{table}[!h]

\caption{\label{tab:Table 1}\textbf{Table 1}. Distribution of retrieved records by source and session. For each source, we reported the number of records, percentage over the session total (after removing duplicates), and the number of records specific for a source as absolute value and as percentage over the source total. All session shows records after joining and deduplication of the Session 1 and Session 2 data set.}
\centering
\resizebox{\linewidth}{!}{
\begin{tabular}[t]{llrlrl}
\toprule
Session & Source & Records & \% over total & Source specific records & \% over source total\\
\midrule
\cellcolor{gray!6}{Session1} & \cellcolor{gray!6}{Total} & \cellcolor{gray!6}{17,755} & \cellcolor{gray!6}{} & \cellcolor{gray!6}{} & \cellcolor{gray!6}{}\\
 & Embase & 12,719 & 71.6\% & 6,683 & 52.5\%\\
\cellcolor{gray!6}{} & \cellcolor{gray!6}{Pubmed} & \cellcolor{gray!6}{9,546} & \cellcolor{gray!6}{53.8\%} & \cellcolor{gray!6}{3,457} & \cellcolor{gray!6}{36.2\%}\\
 & Scopus & 3,175 & 17.9\% & 298 & 9.39\%\\
\cellcolor{gray!6}{} & \cellcolor{gray!6}{WOS} & \cellcolor{gray!6}{2,100} & \cellcolor{gray!6}{11.8\%} & \cellcolor{gray!6}{473} & \cellcolor{gray!6}{22.5\%}\\
\addlinespace
 & IEEE & 60 & 0.34\% & 29 & 48.3\%\\
\cellcolor{gray!6}{Session2} & \cellcolor{gray!6}{Total} & \cellcolor{gray!6}{82,579} & \cellcolor{gray!6}{} & \cellcolor{gray!6}{} & \cellcolor{gray!6}{}\\
 & Embase & 48,396 & 58.6\% & 40,826 & 84.4\%\\
\cellcolor{gray!6}{} & \cellcolor{gray!6}{Pubmed} & \cellcolor{gray!6}{28,811} & \cellcolor{gray!6}{34.9\%} & \cellcolor{gray!6}{18,021} & \cellcolor{gray!6}{62.5\%}\\
 & Scopus & 17,070 & 20.7\% & 4,908 & 28.8\%\\
\addlinespace
\cellcolor{gray!6}{} & \cellcolor{gray!6}{WOS} & \cellcolor{gray!6}{12,956} & \cellcolor{gray!6}{15.7\%} & \cellcolor{gray!6}{2,817} & \cellcolor{gray!6}{21.7\%}\\
 & IEEE & 61 & 0.074\% & 22 & 36.1\%\\
\cellcolor{gray!6}{All Sessions} & \cellcolor{gray!6}{Total} & \cellcolor{gray!6}{98,371} & \cellcolor{gray!6}{} & \cellcolor{gray!6}{} & \cellcolor{gray!6}{}\\
 & Embase & 59,604 & 60.6\% & 46,942 & 78.8\%\\
\cellcolor{gray!6}{} & \cellcolor{gray!6}{Pubmed} & \cellcolor{gray!6}{37,278} & \cellcolor{gray!6}{37.9\%} & \cellcolor{gray!6}{21,371} & \cellcolor{gray!6}{57.3\%}\\
\addlinespace
 & Scopus & 19,353 & 19.7\% & 5,181 & 26.8\%\\
\cellcolor{gray!6}{} & \cellcolor{gray!6}{WOS} & \cellcolor{gray!6}{14,367} & \cellcolor{gray!6}{14.6\%} & \cellcolor{gray!6}{3,175} & \cellcolor{gray!6}{22.1\%}\\
 & IEEE & 108 & 0.11\% & 48 & 44.4\%\\
\bottomrule
\end{tabular}}
\end{table}

The categorised records were used to train the Bayesian classification
model used to label the remaining records. After seven classification
and review (CR) iterations (three resulting in new positive matches and
four extra replications to account for stochastic variability), a total
of 101 positives matches were found, requiring manual review of 766
records (13.2\% positivity rate).\\
It is noticeable how the number of records that required manual review
decreased rapidly between iterations (Table 2), indicating that the
engine was converging while the uncertainties were resolved.\\
This phenomenon is better illustrated in Fig. 1 of Supplemental Material
S2. It shows the mixture distribution of the PPDs of the records,
specifically for records that were manually reviewed, before and after
the classification step: it can be seen how the distribution of
uncertain records shrinks (i.e., it becomes concentrated in a shorter
probability range) and shifts toward the negative zone as more positive
matches are found and reviewed.

\begin{table}[!h]

\caption{\label{tab:Table 2}\textbf{Table 2}. Results of the automatic classification and manual review rounds. The cumulative numbers of positives and negative records and their sum (Total labelled) and percentage over total are shown for each iteration. Also, the number of changes after review and their description is reported."Unlab." indicates unlabelled records marked for review. For each iteration, the number of features used by the engine is also reported. The first row reports the results of the initial manual labelling of records, which served as input for the automatic classification in Iteration 1. In Session 2, the engine uses the labels at the end of Session 1 to classify the newly added records.}
\centering
\resizebox{\linewidth}{!}{
\begin{tabular}[t]{llrrlrrrrrr}
\toprule
Session & Iteration & Positives & Negatives & Total labelled (\%) & Unlab. -> y & Unlab. -> n & Unlab. -> * & n -> y & Changes & N. features\\
\midrule
\cellcolor{gray!6}{Session1 (n = 17755)} & \cellcolor{gray!6}{Initial labelling} & \cellcolor{gray!6}{43} & \cellcolor{gray!6}{207} & \cellcolor{gray!6}{250 (1.41\%)} & \cellcolor{gray!6}{43} & \cellcolor{gray!6}{207} & \cellcolor{gray!6}{0} & \cellcolor{gray!6}{0} & \cellcolor{gray!6}{250} & \cellcolor{gray!6}{2,289}\\
 & 1 & 93 & 529 & 622 (3.5\%) & 50 & 322 & 0 & 0 & 372 & 2,289\\
\cellcolor{gray!6}{} & \cellcolor{gray!6}{2} & \cellcolor{gray!6}{100} & \cellcolor{gray!6}{614} & \cellcolor{gray!6}{714 (4.02\%)} & \cellcolor{gray!6}{6} & \cellcolor{gray!6}{86} & \cellcolor{gray!6}{0} & \cellcolor{gray!6}{1} & \cellcolor{gray!6}{93} & \cellcolor{gray!6}{3,750}\\
 & 3 & 101 & 625 & 726 (4.09\%) & 1 & 11 & 0 & 0 & 12 & 3,834\\
\cellcolor{gray!6}{} & \cellcolor{gray!6}{4} & \cellcolor{gray!6}{101} & \cellcolor{gray!6}{648} & \cellcolor{gray!6}{749 (4.22\%)} & \cellcolor{gray!6}{0} & \cellcolor{gray!6}{23} & \cellcolor{gray!6}{0} & \cellcolor{gray!6}{0} & \cellcolor{gray!6}{23} & \cellcolor{gray!6}{3,856}\\
\addlinespace
 & 5 & 101 & 651 & 752 (4.24\%) & 0 & 3 & 0 & 0 & 3 & 3,856\\
\cellcolor{gray!6}{} & \cellcolor{gray!6}{6} & \cellcolor{gray!6}{101} & \cellcolor{gray!6}{660} & \cellcolor{gray!6}{761 (4.29\%)} & \cellcolor{gray!6}{0} & \cellcolor{gray!6}{9} & \cellcolor{gray!6}{0} & \cellcolor{gray!6}{0} & \cellcolor{gray!6}{9} & \cellcolor{gray!6}{3,856}\\
 & 7 & 101 & 665 & 766 (4.31\%) & 0 & 5 & 0 & 0 & 5 & 3,856\\
\cellcolor{gray!6}{Session2 (n = 98371)} & \cellcolor{gray!6}{1} & \cellcolor{gray!6}{106} & \cellcolor{gray!6}{934} & \cellcolor{gray!6}{1040 (1.06\%)} & \cellcolor{gray!6}{5} & \cellcolor{gray!6}{270} & \cellcolor{gray!6}{998} & \cellcolor{gray!6}{0} & \cellcolor{gray!6}{1,273} & \cellcolor{gray!6}{4,729}\\
 & 2 & 107 & 1,123 & 1230 (1.25\%) & 1 & 189 & 0 & 0 & 190 & 4,729\\
\addlinespace
\cellcolor{gray!6}{} & \cellcolor{gray!6}{3} & \cellcolor{gray!6}{107} & \cellcolor{gray!6}{1,176} & \cellcolor{gray!6}{1283 (1.3\%)} & \cellcolor{gray!6}{0} & \cellcolor{gray!6}{53} & \cellcolor{gray!6}{0} & \cellcolor{gray!6}{0} & \cellcolor{gray!6}{53} & \cellcolor{gray!6}{4,733}\\
 & 4 & 107 & 1,200 & 1307 (1.33\%) & 0 & 24 & 0 & 0 & 24 & 4,729\\
\cellcolor{gray!6}{} & \cellcolor{gray!6}{5} & \cellcolor{gray!6}{107} & \cellcolor{gray!6}{1,209} & \cellcolor{gray!6}{1316 (1.34\%)} & \cellcolor{gray!6}{0} & \cellcolor{gray!6}{9} & \cellcolor{gray!6}{0} & \cellcolor{gray!6}{0} & \cellcolor{gray!6}{9} & \cellcolor{gray!6}{4,729}\\
 & 6 & 107 & 1,226 & 1333 (1.36\%) & 0 & 17 & 0 & 0 & 17 & 4,729\\
\bottomrule
\end{tabular}}
\end{table}

We extracted the 15 more relevant terms for the classification model,
described as: Term (citation part): Inclusion Rate (Inclusion Stability)
{[}linear Relative Risk, Statistic{]}.

Patient Transport (Keyword): 61.2 (3.77) {[}99.1, 21.3{]}, Transfer
(Abstract): 57 (3.93) {[}22.5, 15.4{]}, Network (Title): 56.5 (2.91)
{[}18, 14.2{]}, Network \& Patient (Abstract): 54.2 (4.66) {[}26.3,
15.2{]}, Donker T (Author): 53.5 (4.56) {[}159, 16.5{]}, Worker
(Abstract): 50 (3.33) {[}0.421, -1.21{]}, Hospitals (Keyword): 49.8
(4.31) {[}27.8, 16.5{]}, Movement (Abstract): 47.8 (2.7) {[}27.2, 15{]},
Spread (Title): 46.6 (2.25) {[}16.2, 12.1{]}, Facility (Abstract): 45
(2.22) {[}19.6, 14.8{]}, Orange County (Keyword): 44.3 (3.19) {[}199,
17.2{]}, Conduct (Abstract): 42.6 (3.7) {[}0.221, -2.57{]}, Patient
(Abstract): 42 (3.61) {[}27.6, 7.23{]}, Perform (Abstract): 41.9 (2.38)
{[}0.342, -2.55{]}, Hospital (Title): 39 (1.95) {[}12.5, 12.5{]}.

The ``\&'' indicates nc-ngrams, i.e., terms strongly co-occurrent in the
documents.\\
The engine was able to pick up the central concept of the research
topic, i.e., ``patient transport'' or ``transfer'' through a ``network''
of ``facility''ies that facilitates the ``spread'' of infections, and
even one of the authors of this study (Donker T.) as well as the region
of interest (``Orange County'') of another research group active on the
topic of pathogen spreading over hospital networks. Some terms were
considered highly relevant by the BART models (e.g., ``Worker'' in the
sixth position out of more than 3800 terms considered), although in a
simple linear model, their effect would hardly be significant
(statistic: -1.21 s.e.); these are terms that are only relevant in
conjunction with other terms but not on their own, highlighting the
extra predictive power achieved through the use of advanced, non-linear
machine learning.\\
A more extensive set of terms is presented in Table 1 of Supplemental
Material S2.

\hypertarget{second-session}{%
\subsection{Second session}\label{second-session}}

The results of the first classification session were used to create a
second, data-driven query to perform a more extensive search to find
records that may have been missed during the first search session. The
resulting query was as follows:\\

\emph{(((Donker T) NOT (bacterium isolate)) OR ((network patient) AND
(resistant staphylococcus aureus) NOT (monte carlo) NOT isolation) OR
(facility AND (network patient) AND regional NOT hospitals NOT increase
NOT (patient transport) NOT (control infection use)) OR ((patient
transport) NOT (Donker T) NOT worker) OR (hospitals AND (network
patient) NOT (patient transport) NOT regional NOT clinical) OR (facility
AND (network patient) NOT hospitals NOT (patient transport) NOT regional
NOT prevention NOT medical) OR ((healthcare facility) NOT (Donker T) NOT
worker NOT positive) OR (hospitals NOT (network patient) NOT medical NOT
environmental NOT outcome NOT global) OR ((network patient) NOT facility
NOT hospitals NOT (patient transport) NOT therapy NOT global)) AND
((antimicrobial resistance) OR (healthcare infection))}

The final piece \emph{AND ((antimicrobial resistance) OR (healthcare
infection)} was added manually to define the search domain better since
the algorithm was trained on documents that were all more or less
related to these topics.\\
The generated query also provides a more nuanced understanding of the
engine's internal classification logic, and this is helpful to spot
possible biases in the model.\\

The search was done with the same year filter and procedures used in the
first session.\\

The new search produced 107,294 records (Table 1), of which 48,396
(58.6\%) from the EMBASE, followed by 28,811 (34.9\%) from Pubmed,
17,070 (20.7\%) from SCOPUS, 12,956 (15.7\%) from WOS, and 61 (0.074\%)
from IEEE; compared with the first session, the relative weight of
EMBASE and Pubmed decreased, while the level of content specificity
greatly increased, as it was for SCOPUS. After removal of duplicates,
82,579 unique records were obtained. The newly collected records were
joined with those from the first session and duplicates were removed. We
obtained 98,371 unique records, with just 1,963 shared records between
searches, which equates to 2\% of the total. The percentage of records
shared by two or more sources dropped to 22\%.

Six CR rounds were necessary to complete the second session
classification, with just 6 new positive found after reviewing 568 extra
records. The first CR iteration required the user to review a
substantial number of records (1,273); however, just labelling 275 of
them (the suggested 250 plus 25 already labelled for the framework
hyperparameter tuning) was sufficient to reduce this number to just 190
in the subsequent round. An evaluation of the convergence (Supplemental
Material S2, Fig. 1) showed that, in addition to the dynamics already
observed in session 1 (shrinkage and negative shift), a second mode
appeared in the mixture distribution of the records to be reviewed,
centred in a highly positive zone. The interpretation is that as the
number of negative training records increases, the engine becomes more
and more sceptical and even asks to review some records labelled as
positive in the initial training set generated during Session 1. This
behaviour can rev spot classification errors and inconsistencies.
Considering both sessions, 1,333 records were manually reviewed and 107
(8.03\%) confirmed positive matches were found.

Again, the evaluation of the inclusion rate of the terms showed that the
engine is quite capable of internalising the concepts behind the
research topic. A subsample of the relevant terms used by the model in
the second session is reported in Table 2 of Supplemental Material S2.

\hypertarget{hyperparameter-selection}{%
\subsection{Hyperparameter selection}\label{hyperparameter-selection}}

As described in the methods, hyperparameters were selected by evaluating
sensibility and efficiency through a grid search on a validation set of
1,200 manually labelled records. The analysis suggested that the
following parameter combination performed best: an initial training set
of 250 categorised records with 10x oversampling of positive matches,
ten models in the ensemble, no bootstrapping and an uncertainty zone
defined by the 98\% predictive interval. This combination of parameters
was associated with a sensitivity of 98.8\% (81 / 82 positive matches
found) and an efficiency of 61.5\% (462 / 1200 records evaluated). The
detailed results of the hyperparameter tuning analysis are reported in
Table 3 of Supplemental Material S2. Fig. 2 in Supplemental Material S2
demonstrates that the positive record oversampling rate, the number of
ensemble models and the size of the initial training set were the
parameters that mainly impacted performance.

\hypertarget{performance-evaluation-1}{%
\subsection{Performance evaluation}\label{performance-evaluation-1}}

To evaluate the theoretical performance of the engine, a surrogate
Bayesian logistic regression model was trained on the manually reviewed
labels using only the lower boundary of the record PPDs as predictor
(see the Methods for details). The surrogate model showed the high
predictive power of the scores produced by the classification model
(Bayesian R2: 98.1\% {[}97.4\%, 98.3\%{]} for session 1 and 98.2\%
{[}97.6\%, 98.3\%{]} for session 2).\\

Fig. 2 presents the actual and predicted (from the surrogate model)
cumulative number of positive matches, ordered by the initial simple
ordering query: the median of surrogate models' cumulative predictive
distributions matches the actual number of positive records found quite
well. It is striking how many more records would have required manual
evaluation to find the same number of positive matches without a
classification algorithm, with some positive matches found close to the
end of the heuristically ordered list of records.\\

Table 3 shows various performance indexes for both sessions, both
descriptive (Total records, Reviewed records, Observed positive matches)
and estimated through the surrogate model (Expected efficiency,
Predicted positive matches, Expected sensitivity, \(R^2\)).\\
In session 1 we observe an expected total number of positives of 101
{[}101, 108{]} for an estimated sensitivity of 100\% {[}93.5\%, 100\%{]}
and efficiency of 95.6\% {[}95.3\%, 95.7\%{]}. In session 2 we observed
a drop in expected sensitivity, especially in the lower credibility
boundary (97.3\% {[}72.8\%, 100\%{]}): as the number of records
increases, even a small probability of being a positive match can, in
the worst-case scenario, lead to a relevant number of predicted positive
matches (147 in this case). To ensure no obvious positive matches were
missed, we evaluated 100 non-reviewed records with the highest median
predicted probability and found no additional positive matches.\\

\begin{table}[!h]

\caption{\label{tab:Table 3}\textbf{Table 3}. Estimated performance summary. The table reports for each session, the number of reviewed records and the percentage over the total. Also, the posterior expected number of positive records, sensitivity and efficiency (as WSoR) are reported, with their 90\% PrI truncated to the observed realisation in the dataset [trunc. PrI] (see. methods). Finally, the logistic model's median Bayesian $R^2$ [90\% CrI] is reported. PrI: Predictive Intervals; CrI: Credibility Intervals.}
\centering
\begin{tabular}[t]{lll}
\toprule
Indicator & Session 1 & Session 2\\
\midrule
\cellcolor{gray!6}{Total records} & \cellcolor{gray!6}{17,755} & \cellcolor{gray!6}{98,371}\\
Reviewed records (\% over total records) & 766 (4.31\%) & 1,333 (1.36\%)\\
\cellcolor{gray!6}{Expected efficiency (over random) [trunc. 90\% PrI]} & \cellcolor{gray!6}{95.6\% [95.3\%, 95.7\%]} & \cellcolor{gray!6}{98.6\% [98.1\%, 98.6\%]}\\
Observed positive matches (\% over total records) & 101 (0.57\%) & 107 (0.11\%)\\
\cellcolor{gray!6}{Predicted positive matches [trunc. 90\% PrI]} & \cellcolor{gray!6}{101 [101, 108]} & \cellcolor{gray!6}{110 [107, 147]}\\
\addlinespace
Expected sensitivity [trunc. 90\% PrI] & 100\% [93.5\%, 100\%] & 97.3\% [72.8\%, 100\%]\\
\cellcolor{gray!6}{Simple Model $R^2$ [90\% CrI]} & \cellcolor{gray!6}{98.1\% [97.4\%, 98.3\%]} & \cellcolor{gray!6}{98.2\% [97.6\%, 98.3\%]}\\
\bottomrule
\end{tabular}
\end{table}

\begin{figure}
\centering
\includegraphics{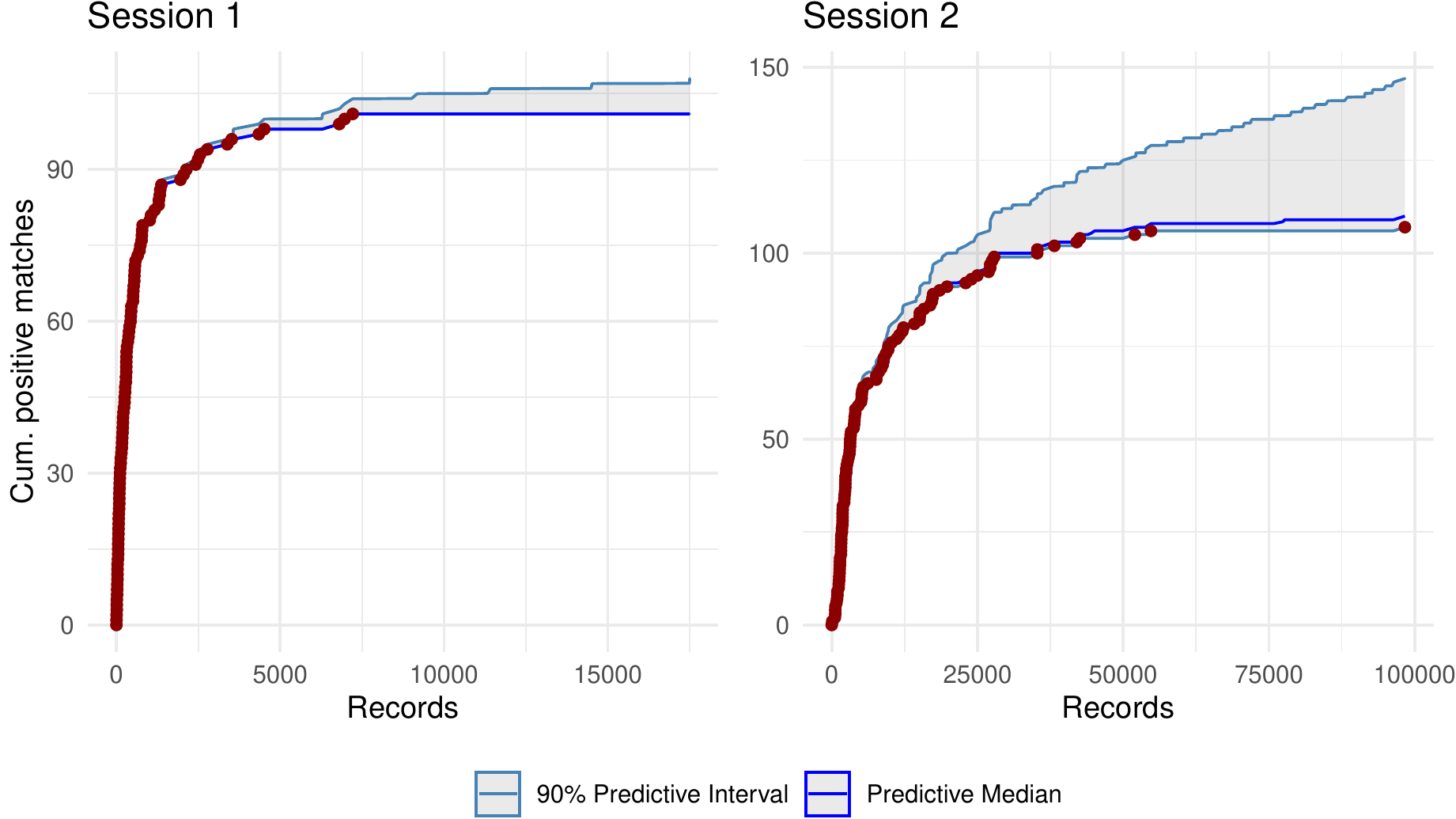}
\caption{\textbf{Figure 2}. Observed cumulative number of positive
matches (red dots) sorted by simple query ordering. The {[}trunc. 90\%
PrI{]} of the cumulative positive matches estimated by the Bayesian
logistic model is shown as a shaded area delimited by the 95\% quantile
of the PrI and by the observed number of positive matches (light blue
lines). A darker blue line represents the median of the PrI.}
\end{figure}

\hypertarget{discussion}{%
\section{Discussion}\label{discussion}}

We propose a new integrated framework to help researchers collect and
screen scientific publications characterised by high performance and
versatility. This framework joins the joining the growing field of
systematic review automation (SRA) and helpers (SRH) tools (A. M. Cohen
et al. 2006, 2010; Ananiadou et al. 2009; O'Mara-Eves et al. 2015). This
framework implements standard approaches and uses ad-hoc solutions to
common SRA issues. By freely sharing the tool as an open-source R
package and by following a modular design, we sought to adopt some of
the so-called Vienna Principles advocated by the International
Collaboration for the Automation of Systematic Reviews (ICASR) (E.
Beller et al. 2018).\\
The framework consists of four main components: 1) an integrated
query-based citation search and management engine, 2) a Bayesian active
machine learning-based citation classifier, and 3) a data-driven search
query generation algorithm.\\

The search engine module used by the framework can automatically collect
citation data from three well-known scientific databases (i.e., Pubmed,
Web of Science, and the Institute of Electrical and Electronics
Engineers) and process manually downloaded results from two more sources
(SCOPUS, EMBASE). In comparison, most commercial or free SRH tools rely
on internal databases (e.g., Mendeley \url{https://www.mendeley.com/})
sometimes focusing only on a particular topic (Visser 2010) or a single
external data source (Thomas and Brunton 2007; Poulter et al. 2008;
Soto, Przybyła, and Ananiadou 2019).\\
Mixing different databases is essential to obtain a more comprehensive
view of the literature (Bajpai et al. 2011; Wilkins, Gillies, and Davies
2005; Woods and Trewheellar 1998): in our results, 18.7\% of the
positive matches were found in only one of the different data sources,
and no positive record was present in all the sources (data not
shown).\\
The framework online search algorithms are efficient enough to manage
tens of thousands of search results, using various solutions to overcome
the limitations of citation databases in terms of traffic and download
quotas. The results are then automatically organised, deduplicated and
arranged by ``simple query ordering'' in a uniform corpus. The
preliminary ordering increases the positivity rate in the initial
training set (Wallace, Small, et al. 2010).\\

For the framework's record screening module, we developed an active
machine learning protocol (Settles 2009; Miwa et al. 2014) based on the
best practices from other SRA studies, bringing further improvements at
various levels.\\
The feature extractor module uses modern NLP techniques (Ananiadou and
McNaught 2006; K. B. Cohen and Hunter 2008) to transform free text into
input data for machine learning. We did not include classical n-grams
(Schonlau and Guenther 2017); rather, we used network analysis to find
non-consecutive, frequently associated terms, a generalisation of
n-grams that relaxes the term adjacency assumption. This approach can
also incorporate term connections across different parts of the records,
e.g., terms having a different relevance when associated with a
particular author. The same technique was used with different parameters
to merge redundant terms, increasing estimation efficiency and reducing
noise.\\
The use of concurrency network-driven text modelling is not new
(Francois Rousseau 2015; Violos et al. 2016; François Rousseau, Kiagias,
and Vazirgiannis 2015; Ohsawa, Benson, and Yachida 1998) and is a
valuable tool to extract semantic information that is not evident in
one-word or consecutive n-gram models.\\

The automatic classification algorithm is based on Bayesian Additive
Regression Trees (BART) (Chipman et al. 2010; Kapelner and Bleich 2013).
Like other boosted trees algorithms (Hastie, Tibshirani, and Friedman
2009), the BART method can explore complex non-linearities, perform
variable selection, manage missing data while maintaining high
predictive power.\\
However, the Bayesian foundation of the method provides further
benefits: lower sensitivity to the choice of hyperparameters, natural
regularisation through priors, and, most importantly, predictive
distributions as output instead of point-wise predictions (Soria-Olivas
et al. 2011; Joo, Chung, and Seo 2020; Jospin et al. 2020). By selecting
relatively tight prior distributions, we discouraged overly deep trees,
long tree sequences, and extreme predicted probabilities, thus reducing
the risk of overfitting.\\
The algorithm runs multiple replications of the model and averages their
predictive distributions creating an ``ensemble''; this technique has
been shown to improve out-of-sample predictive performance (Zhou 2021;
Dietterich 2000), as confirmed during the hyperparameter evaluation
(Supplemental Material S2). Ensembling reduces the uncertainty in the
predictive distribution tails related to the randomness in the MCMC fit
(Robert and Casella 2004), generating a shift in the probability mass
towards the distribution centre and stabilising it (i.e., reducing
variance without impacting bias). On the other hand, simply imposing
more robust uninformative priors against extreme predictions would have
reduced variance but also shifted the distribution towards a
non-decision zone, increasing bias (Hansen et al. 2000).\\
Since the number of model replications has a significant impact on
computation times, we decided to use ten replicas, the lower value after
which performance stabilised, as resulted from the evaluation of the
hyperparameters (Supplemental Material S2, Fig. 2).\\
We also investigated whether bootstrapping between replications (Breiman
1996) would improve performance; however, contrary to theory
(Díez-Pastor et al. 2015), it appeared to be slightly detrimental in our
case (Supplemental Material S2, Fig. 2) compared to simple ensembling.\\

A low proportion of relevant matches (class imbalance) is typical for
literature reviews (Sampson, Tetzlaff, and Urquhart 2011; Wallace,
Trikalinos, et al. 2010; O'Mara-Eves et al. 2015), and a strong
imbalance between positive and negative records can affect sensitivity
(Khoshgoftaar, Van Hulse, and Napolitano 2010; Chawla, Japkowicz, and
Kotcz 2004).\\
To overcome this problem, we oversampled (Batista, Prati, and Monard
2004) the positive records ten times before model fitting. The
hyperparameter analysis showed that the oversampling rate, together with
model ensembling, was the parameter with the most significant impact on
performance.\\
A known risk with positive oversampling is the misclassification of
negative records (Ramezankhani et al. 2016). However, since all
predicted positives in our approach are reviewed manually, we are always
guaranteed to achieve 100\% specificity/positive predictive value: the
only price for the increased sensitivity due to oversampling is a larger
number of records to be reviewed.\\
An alternative to oversampling would be to apply different weights
and/or costs to the classes (Abd Elrahman and Abraham 2013; Díez-Pastor
et al. 2015), but the BART implementation we used did not have this
feature; furthermore, using simple oversampling allows for a broader
compatibility with different modelling engines (Galar et al. 2011;
Roshan and Asadi 2020).\\
Finally, sorting the records by query term frequency (simple query
ordering) produces a much higher rate of relevant records in the initial
training set (17.2\%) compared to the overall data (0.11\%), which
boosts the sensitivity of the model.\\

One of the key innovations we have introduced is the concept of
``uncertainty zone'', the implementation of which is possible thanks to
the Bayesian foundation of the classification model.\\
This construct guides the selection of records to be manually reviewed
and gets dynamically updated and reduced after each CR iteration, as
more uncertain predictions are evaluated (Supplemental Material S2 Fig.
1).\\
The use of a dynamic uncertainty zone overcomes the usual requirement of
dataset-specific hard thresholds in active machine learning and allows
to review multiple items at once between iterations (Laws and Schütze
2008; Miwa et al. 2014; Zhu et al. 2010). The hyperparameters required
by our algorithm are general and non-task-specific, like the PPD
intervals underlying the uncertainty zone and the maximum number of
iterations without positive matches after which a session is concluded;
the evaluation of the classification model hyperparameters shows that
the algorithm is robust against variations in these parameters, and we
expect the default values to perform well on most datasets.\\
Since researchers are asked to review both records predicted as surely
relevant and those inside the uncertainty zone, this method can be
considered as a unifying synthesis of the ``certainty'' and
``uncertainty'' paradigms of active learning (Miwa et al. 2014).\\

We assessed performance as the ability of the screening procedure
(automatic classification plus manual review) to find the largest number
of relevant records while requiring manual reviewing for as few of them
as possible (i.e., sensitivity \(\times\) efficiency).\\
We avoided the classical out-of-sample approaches such as train-test
sampling, out-of-bag bootstrapping or cross-validation (Kohavi et al.
1995; James et al. 2013). Such methods primarily assume that the rate of
positivity is the same on average in every possible random subset of the
data (Tashman 2000); this uniformity is broken by how the initial
training set and the subsequent reviewed records are selected by the
query-based ordering and active learning algorithm, resulting in a lower
positivity rate in the unlabelled records (Fig. 2). Moreover, a
literature corpus is unique per search query/database combination, and
therefore any out-of-sample performance estimate is not replicable since
no new data can be acquired related to the current corpus.\\
To estimate overall sensitivity, we instead applied simple Bayesian
regression (surrogate model) to the manually reviewed data to abstract
the classification model predictions and generate a maximum entropy
(Harremoës and Topsøe 2001) estimate of the number of missed positive
matches among the unreviewed records in the whole dataset. This simple
surrogate model fitted the data very well (\(R^2\) consistently above
97\%) using only the lower 98\% PrI boundary of the PPDs as predictor,
indicating predictive consistency in the classification model. The
posterior predictive distribution of the surrogate model could be used
to explore worse case scenarios in terms of sensitivity.\\

Our framework achieves very high sensitivity by screening only a very
small fraction of all records, bringing a meaningful reduction in
workload.\\
Based on the surrogate model, we predicted a predicted median
sensitivity of 100\% {[}93.5\%, 100\%{]} in the first session (screening
4.29\% of records) and of 97.3\% {[}73.8\%, 100\%{]} in the second
(screening 1.34\% of records): efficiency increased significantly in the
second session as only a few new positive matches were found; however,
given the large number of records, uncertainty about sensitivity
increased, as expected.\\
Both results are above the usual performance in this field (O'Mara-Eves
et al. 2015) and are in line with the average sensitivity of 92\%
estimated after human-only screening (Edwards et al. 2002). In one
interesting case, the model detected a human-caused misclassification
error, demonstrating its robustness and value as a second screener, a
role already suggested for SRA tools in previous studies (Frunza,
Inkpen, and Matwin 2010; Bekhuis and Demner-Fushman 2012, 2010).
Although ``simple query ordering'' concentrated most relevant matches in
the first 20-25 thousand records, without the tool support, the
remaining relevant records would have been missed without manually
screening almost the entire dataset.\\

The model required \textasciitilde5-20 minutes per iteration to perform
the predictions in session 1 (17,755 documents) and 20-40 minutes in
session 2 (98,371 documents) on an eight-core, 2.5 GHz, 16 GB RAM, 2014
laptop; including manual record review, one session required 1-3 days of
work, for a total of 1-2 weeks for the whole process (including record
collection). This is a considerable time saving compared to the several
months typically required for the screening phase of systematic reviews
(Bannach-Brown et al. 2019; Borah et al. 2017; Allen and Olkin 1999). To
our knowledge, the amount of data processed (\textasciitilde100,000
records) was larger than what is typical of most SRA studies
(O'Mara-Eves et al. 2015; Olorisade et al. 2016), highlighting the
scalability of the tool in real-world scenarios.\\

The last module of our framework is an algorithm for data-driven search
query generation. Generating an efficient and effective search query is
a complex task (Lefebvre et al. 2011; Hammerstrøm et al. 2010); it
requires building a combination of positive and negative terms to
maximise the number of relevant search results while minimising the
total number of records to be reviewed. Our solution combines a
sensitivity-driven subquery proposal engine based on concurrent decision
trees (Blanco-Justicia and Domingo-Ferrer 2019; Moore et al. 2018) built
on the BART ensemble PPD, with a human review step and an
efficiency-driven query builder. The aim is to generate a new search
query to help find records missed in the first search session. The
generated query did indeed retrieve a few more relevant records not
found in session 1 but at the cost of significantly increasing the
number of documents.\\
An interesting aspect of this feature is that it provides a
human-readable overview of the classification rules learned by the
classification model, showing which combination of terms was
particularly relevant and even spotting authors and geographical
locations associated with the study topic. The generated query,
therefore, served also as a means for machine learning explainability
(Bhatt et al. 2020; Burkart and Huber 2021), useful for understanding
and detecting biases in black-box classification algorithms (Malhi,
Knapic, and Främling 2020); explainability is often required or even
legally mandatory for high-stake machine learning applications (Bibal et
al. 2021, 2020).\\
It is important to note that this process is entirely data-driven. The
algorithm is only aware of the ``world'' defined by the dataset used as
input, which is generated by a specific search query focused on a
particular topic. Therefore, the new query may not be specific enough
when applied to an unbounded search domain and may return an
unmanageable amount of irrelevant results. The solution we found was to
add another component to the query, specifying the general topic
(antimicrobial resistance and healthcare-associated infections) of our
research.\\

As mentioned early, our framework builds on modularity. We have designed
so that each module can become fully independent in future iterations;
it will be possible for users to add custom features such as citation
search and parsing for other scientific databases, alternative text
processing algorithms or machine learning modules. We consider such
interoperability to be extremely relevant: the main strength of our tool
lies in the composition of many independent solutions, such as the idea
of Bayesian active machine learning and the exploit of the derived
uncertainty in defining the records needing human review.\\
Each component could benefit considerably from the recent improvements
in text mining and machine learning.\\
For example, the text processing approach based on the ``boolean
bag-of-words'' paradigm is quite simple and could be improved by more
nuanced text representations. It might be considered whether feature
transformations such as TF-IDF (Baeza-Yates, Ribeiro-Neto, et al. 1999;
Ananiadou and McNaught 2006) could be advantageous, although we
hypothesise that tree-based classification algorithms like BART are
robust enough not to require such operations. Instead, it might be worth
exploring the application of word embedding: this technique transforms
terms into semantic vectors derived from the surrounding text (Turian,
Ratinov, and Bengio 2010; Bollegala, Maehara, and Kawarabayashi 2015;
Minaee et al. 2021) and could be used to reduce noise by merging
different terms that are semantically similar or enhance signal by
distinguishing identical terms with different meaning given the context.
Another option would be to employ unsupervised learning models like
Latent Dirichlet Analysis and Latent Semantic Analysis, (Pavlinek and
Podgorelec 2017; Q. Chen, Yao, and Yang 2016; Landauer, Foltz, and Laham
1998) or graph-of-word techniques (Ohsawa, Benson, and Yachida 1998;
Francois Rousseau 2015) to extract topics that expand the feature
space.\\
Our classification algorithm is applicable with any Bayesian supervised
machine learning method that provides full PPDs; therefore, alternative
classification models, such as Gaussian Processes, known for their
flexibility (Jayashree and Srijith 2020; S.-H. Chen et al. 2015), could
be evaluated. It would be even more interesting to test advanced
learning algorithms that go beyond the bag-of-words approach and take
into consideration higher-level features in the text such as term
context and sequences, long-distance term relationships, semantic
structures, etc., (Cheng et al. 2019; Minaee et al. 2021; Li et al.
2020; J. Yang, Bai, and Guo 2020; Lai et al. 2015; Farkas 1995),
provided that a Bayesian implementation of such algorithms is available
(for example C. Chen, Lin, and Terejanu (2018)).\\
Finally, a natural improvement would be to provide a graphical user
interface to make the framework easy to use also for less technical
users.

The field of literature review automation is evolving rapidly, and we
anticipate an increasing use of such technologies to address the
accelerating pace of scientific production. We believe it is encouraging
that a wide variety of tools are being made available to let researchers
and policymakers find the approach that best fits their needs.\\
We contribute to this field with an innovative framework that provides
excellent performance and easy integration with existing systematic
review pipelines. The value of this work lies not only in the framework
itself, which we make available as open-source software, but also in the
set of methodologies we developed to solve various SRA issues and which
can also be used to improve existing solutions.\\

\newpage

\hypertarget{references}{%
\section*{References}\label{references}}
\addcontentsline{toc}{section}{References}

\hypertarget{refs}{}
\begin{CSLReferences}{1}{0}
\leavevmode\vadjust pre{\hypertarget{ref-abd2013review}{}}%
Abd Elrahman, Shaza M, and Ajith Abraham. 2013. {`A Review of Class
Imbalance Problem'}. \emph{Journal of Network and Innovative Computing}
1 (2013): 332--40.

\leavevmode\vadjust pre{\hypertarget{ref-allen1999estimating}{}}%
Allen, I Elaine, and Ingram Olkin. 1999. {`Estimating Time to Conduct a
Meta-Analysis from Number of Citations Retrieved'}. \emph{Jama} 282 (7):
634--35.

\leavevmode\vadjust pre{\hypertarget{ref-ananiadou2006text}{}}%
Ananiadou, Sophia, and John McNaught. 2006. \emph{Text Mining for
Biology and Biomedicine}. Citeseer.

\leavevmode\vadjust pre{\hypertarget{ref-ananiadou2009supporting}{}}%
Ananiadou, Sophia, Brian Rea, Naoaki Okazaki, Rob Procter, and James
Thomas. 2009. {`Supporting Systematic Reviews Using Text Mining'}.
\emph{Social Science Computer Review} 27 (4): 509--23.

\leavevmode\vadjust pre{\hypertarget{ref-aviv2021publication}{}}%
Aviv-Reuven, Shir, and Ariel Rosenfeld. 2021. {`Publication Patterns'
Changes Due to the COVID-19 Pandemic: A Longitudinal and Short-Term
Scientometric Analysis'}. \emph{Scientometrics}, 1--24.

\leavevmode\vadjust pre{\hypertarget{ref-babar2009systematic}{}}%
Babar, Muhammad Ali, and He Zhang. 2009. {`Systematic Literature Reviews
in Software Engineering: Preliminary Results from Interviews with
Researchers'}. In \emph{2009 3rd International Symposium on Empirical
Software Engineering and Measurement}, 346--55. IEEE.

\leavevmode\vadjust pre{\hypertarget{ref-baeza1999modern}{}}%
Baeza-Yates, Ricardo, Berthier Ribeiro-Neto, et al. 1999. \emph{Modern
Information Retrieval}. Vol. 463. ACM press New York.

\leavevmode\vadjust pre{\hypertarget{ref-bajpai2011search}{}}%
Bajpai, Akhilesh, Sravanthi Davuluri, Haritha Haridas, Greta Kasliwal, H
Deepti, KS Sreelakshmi, Darshan Chandrashekar, et al. 2011. {`In Search
of the Right Literature Search Engine (s)'}. \emph{Nature Precedings},
1--1.

\leavevmode\vadjust pre{\hypertarget{ref-bannach2019machine}{}}%
Bannach-Brown, Alexandra, Piotr Przybyła, James Thomas, Andrew SC Rice,
Sophia Ananiadou, Jing Liao, and Malcolm Robert Macleod. 2019. {`Machine
Learning Algorithms for Systematic Review: Reducing Workload in a
Preclinical Review of Animal Studies and Reducing Human Screening
Error'}. \emph{Systematic Reviews} 8 (1): 1--12.

\leavevmode\vadjust pre{\hypertarget{ref-bastian2010seventy}{}}%
Bastian, Hilda, Paul Glasziou, and Iain Chalmers. 2010. {`Seventy-Five
Trials and Eleven Systematic Reviews a Day: How Will We Ever Keep Up?'}.
\emph{PLoS Medicine} 7 (9): e1000326.

\leavevmode\vadjust pre{\hypertarget{ref-batista2004study}{}}%
Batista, Gustavo EAPA, Ronaldo C Prati, and Maria Carolina Monard. 2004.
{`A Study of the Behavior of Several Methods for Balancing Machine
Learning Training Data'}. \emph{ACM SIGKDD Explorations Newsletter} 6
(1): 20--29.

\leavevmode\vadjust pre{\hypertarget{ref-bekhuis2010towards}{}}%
Bekhuis, Tanja, and Dina Demner-Fushman. 2010. {`Towards Automating the
Initial Screening Phase of a Systematic Review'}. \emph{MEDINFO 2010},
146--50.

\leavevmode\vadjust pre{\hypertarget{ref-bekhuis2012screening}{}}%
---------. 2012. {`Screening Nonrandomized Studies for Medical
Systematic Reviews: A Comparative Study of Classifiers'}.
\emph{Artificial Intelligence in Medicine} 55 (3): 197--207.

\leavevmode\vadjust pre{\hypertarget{ref-beller2013systematic}{}}%
Beller, Elaine M, Joyce Kee-Hsin Chen, Una Li-Hsiang Wang, and Paul P
Glasziou. 2013. {`Are Systematic Reviews up-to-Date at the Time of
Publication?'}. \emph{Systematic Reviews} 2 (1): 1--6.

\leavevmode\vadjust pre{\hypertarget{ref-beller2018making}{}}%
Beller, Elaine, Justin Clark, Guy Tsafnat, Clive Adams, Heinz Diehl,
Hans Lund, Mourad Ouzzani, et al. 2018. {`Making Progress with the
Automation of Systematic Reviews: Principles of the International
Collaboration for the Automation of Systematic Reviews (ICASR)'}.
\emph{Systematic Reviews} 7 (1): 1--7.

\leavevmode\vadjust pre{\hypertarget{ref-bhatt2020machine}{}}%
Bhatt, Umang, McKane Andrus, Adrian Weller, and Alice Xiang. 2020.
{`Machine Learning Explainability for External Stakeholders'}.
\emph{arXiv Preprint arXiv:2007.05408}.

\leavevmode\vadjust pre{\hypertarget{ref-bibal2021legal}{}}%
Bibal, Adrien, Michael Lognoul, Alexandre De Streel, and Benoît Frénay.
2021. {`Legal Requirements on Explainability in Machine Learning'}.
\emph{Artificial Intelligence and Law} 29 (2): 149--69.

\leavevmode\vadjust pre{\hypertarget{ref-bibal2020impact}{}}%
Bibal, Adrien, Michael Lognoul, Alexandre de Streel, and Benoît Frénay.
2020. {`Impact of Legal Requirements on Explainability in Machine
Learning'}. \emph{arXiv Preprint arXiv:2007.05479}.

\leavevmode\vadjust pre{\hypertarget{ref-blanco2019machine}{}}%
Blanco-Justicia, Alberto, and Josep Domingo-Ferrer. 2019. {`Machine
Learning Explainability Through Comprehensible Decision Trees'}. In
\emph{International Cross-Domain Conference for Machine Learning and
Knowledge Extraction}, 15--26. Springer.

\leavevmode\vadjust pre{\hypertarget{ref-bollegala2015embedding}{}}%
Bollegala, Danushka, Takanori Maehara, and Ken-ichi Kawarabayashi. 2015.
{`Embedding Semantic Relations into Word Representations'}. In
\emph{Twenty-Fourth International Joint Conference on Artificial
Intelligence}.

\leavevmode\vadjust pre{\hypertarget{ref-borah2017analysis}{}}%
Borah, Rohit, Andrew W Brown, Patrice L Capers, and Kathryn A Kaiser.
2017. {`Analysis of the Time and Workers Needed to Conduct Systematic
Reviews of Medical Interventions Using Data from the PROSPERO
Registry'}. \emph{BMJ Open} 7 (2): e012545.

\leavevmode\vadjust pre{\hypertarget{ref-bornmann2015growth}{}}%
Bornmann, Lutz, and Rüdiger Mutz. 2015. {`Growth Rates of Modern
Science: A Bibliometric Analysis Based on the Number of Publications and
Cited References'}. \emph{Journal of the Association for Information
Science and Technology} 66 (11): 2215--22.

\leavevmode\vadjust pre{\hypertarget{ref-breiman1996bagging}{}}%
Breiman, Leo. 1996. {`Bagging Predictors'}. \emph{Machine Learning} 24
(2): 123--40.

\leavevmode\vadjust pre{\hypertarget{ref-burkart2021survey}{}}%
Burkart, Nadia, and Marco F Huber. 2021. {`A Survey on the
Explainability of Supervised Machine Learning'}. \emph{Journal of
Artificial Intelligence Research} 70: 245--317.

\leavevmode\vadjust pre{\hypertarget{ref-chae1993presenting}{}}%
Chae, Kyung-Chul. 1993. {`Presenting the Negative Hypergeometric
Distribution to the Introductory Statistics Courses'}.
\emph{International Journal of Mathematical Education in Science and
Technology} 24 (4): 523--26.

\leavevmode\vadjust pre{\hypertarget{ref-chawla2004special}{}}%
Chawla, Nitesh V, Nathalie Japkowicz, and Aleksander Kotcz. 2004.
{`Special Issue on Learning from Imbalanced Data Sets'}. \emph{ACM
SIGKDD Explorations Newsletter} 6 (1): 1--6.

\leavevmode\vadjust pre{\hypertarget{ref-chen2018approximate}{}}%
Chen, Chao, Xiao Lin, and Gabriel Terejanu. 2018. {`An Approximate
Bayesian Long Short-Term Memory Algorithm for Outlier Detection'}. In
\emph{2018 24th International Conference on Pattern Recognition (ICPR)},
201--6. IEEE.

\leavevmode\vadjust pre{\hypertarget{ref-chen2016short}{}}%
Chen, Qiuxing, Lixiu Yao, and Jie Yang. 2016. {`Short Text
Classification Based on LDA Topic Model'}. In \emph{2016 International
Conference on Audio, Language and Image Processing (ICALIP)}, 749--53.
IEEE.

\leavevmode\vadjust pre{\hypertarget{ref-chen2015gaussian}{}}%
Chen, Sih-Huei, Yuan-Shan Lee, Tzu-Chiang Tai, and Jia-Ching Wang. 2015.
{`Gaussian Process Based Text Categorization for Healthy Information'}.
In \emph{2015 International Conference on Orange Technologies (ICOT)},
30--33. \url{https://doi.org/10.1109/ICOT.2015.7498487}.

\leavevmode\vadjust pre{\hypertarget{ref-cheng2019document}{}}%
Cheng, Y, Z Ye, M Wang, and Q Zhang. 2019. {`Document Classification
Based on Convolutional Neural Network and Hierarchical Attention
Network'}. \emph{Neural Network World} 29 (2): 83--98.

\leavevmode\vadjust pre{\hypertarget{ref-chipman2010bart}{}}%
Chipman, Hugh A, Edward I George, Robert E McCulloch, et al. 2010.
{`BART: Bayesian Additive Regression Trees'}. \emph{The Annals of
Applied Statistics} 4 (1): 266--98.

\leavevmode\vadjust pre{\hypertarget{ref-claesen2015hyperparameter}{}}%
Claesen, Marc, and Bart De Moor. 2015. {`Hyperparameter Search in
Machine Learning'}. \emph{arXiv Preprint arXiv:1502.02127}.

\leavevmode\vadjust pre{\hypertarget{ref-cohen2010evidence}{}}%
Cohen, Aaron M, Clive E Adams, John M Davis, Clement Yu, Philip S Yu,
Weiyi Meng, Lorna Duggan, Marian McDonagh, and Neil R Smalheiser. 2010.
{`Evidence-Based Medicine, the Essential Role of Systematic Reviews, and
the Need for Automated Text Mining Tools'}. In \emph{Proceedings of the
1st ACM International Health Informatics Symposium}, 376--80.

\leavevmode\vadjust pre{\hypertarget{ref-cohen2006reducing}{}}%
Cohen, Aaron M, William R Hersh, Kim Peterson, and Po-Yin Yen. 2006.
{`Reducing Workload in Systematic Review Preparation Using Automated
Citation Classification'}. \emph{Journal of the American Medical
Informatics Association} 13 (2): 206--19.

\leavevmode\vadjust pre{\hypertarget{ref-cohen2008getting}{}}%
Cohen, K Bretonnel, and Lawrence Hunter. 2008. {`Getting Started in Text
Mining'}. \emph{PLoS Computational Biology} 4 (1): e20.

\leavevmode\vadjust pre{\hypertarget{ref-dietterich2000ensemble}{}}%
Dietterich, Thomas G. 2000. {`Ensemble Methods in Machine Learning'}. In
\emph{International Workshop on Multiple Classifier Systems}, 1--15.
Springer.

\leavevmode\vadjust pre{\hypertarget{ref-diez2015diversity}{}}%
Díez-Pastor, José F, Juan J Rodríguez, César I García-Osorio, and
Ludmila I Kuncheva. 2015. {`Diversity Techniques Improve the Performance
of the Best Imbalance Learning Ensembles'}. \emph{Information Sciences}
325: 98--117.

\leavevmode\vadjust pre{\hypertarget{ref-edwards2002identification}{}}%
Edwards, Phil, Mike Clarke, Carolyn DiGuiseppi, Sarah Pratap, Ian
Roberts, and Reinhard Wentz. 2002. {`Identification of Randomized
Controlled Trials in Systematic Reviews: Accuracy and Reliability of
Screening Records'}. \emph{Statistics in Medicine} 21 (11): 1635--40.

\leavevmode\vadjust pre{\hypertarget{ref-eppstein2010listing}{}}%
Eppstein, David, Maarten Löffler, and Darren Strash. 2010. {`Listing All
Maximal Cliques in Sparse Graphs in Near-Optimal Time'}. In
\emph{International Symposium on Algorithms and Computation}, 403--14.
Springer.

\leavevmode\vadjust pre{\hypertarget{ref-farkas1995document}{}}%
Farkas, Jennifer. 1995. {`Document Classification and Recurrent Neural
Networks'}. In \emph{Proceedings of the 1995 Conference of the Centre
for Advanced Studies on Collaborative Research}, 21.

\leavevmode\vadjust pre{\hypertarget{ref-frunza2010building}{}}%
Frunza, Oana, Diana Inkpen, and Stan Matwin. 2010. {`Building Systematic
Reviews Using Automatic Text Classification Techniques'}. In
\emph{Coling 2010: Posters}, 303--11.

\leavevmode\vadjust pre{\hypertarget{ref-galar2011review}{}}%
Galar, Mikel, Alberto Fernandez, Edurne Barrenechea, Humberto Bustince,
and Francisco Herrera. 2011. {`A Review on Ensembles for the Class
Imbalance Problem: Bagging-, Boosting-, and Hybrid-Based Approaches'}.
\emph{IEEE Transactions on Systems, Man, and Cybernetics, Part C
(Applications and Reviews)} 42 (4): 463--84.

\leavevmode\vadjust pre{\hypertarget{ref-gelman2019r}{}}%
Gelman, Andrew, Ben Goodrich, Jonah Gabry, and Aki Vehtari. 2019.
{`R-Squared for Bayesian Regression Models'}. \emph{The American
Statistician}.

\leavevmode\vadjust pre{\hypertarget{ref-hammerstrom2010searching}{}}%
Hammerstrøm, Karianne, Anne Wade, Anne-Marie Klint Jørgensen, and
Karianne Hammerstrøm. 2010. {`Searching for Studies'}. \emph{Education}
54 (11.3).

\leavevmode\vadjust pre{\hypertarget{ref-hansen2000bayesian}{}}%
Hansen, Lars Kai et al. 2000. {`Bayesian Averaging Is Well-Temperated'}.
In \emph{Proceedings of NIPS}, 99:265--71.

\leavevmode\vadjust pre{\hypertarget{ref-harremoes2001maximum}{}}%
Harremoës, Peter, and Flemming Topsøe. 2001. {`Maximum Entropy
Fundamentals'}. \emph{Entropy} 3 (3): 191--226.

\leavevmode\vadjust pre{\hypertarget{ref-hastie2009boosting}{}}%
Hastie, Trevor, Robert Tibshirani, and Jerome Friedman. 2009. {`Boosting
and Additive Trees'}. In \emph{The Elements of Statistical Learning},
337--87. Springer.

\leavevmode\vadjust pre{\hypertarget{ref-higgins2019cochrane}{}}%
Higgins, Julian PT, James Thomas, Jacqueline Chandler, Miranda Cumpston,
Tianjing Li, Matthew J Page, and Vivian A Welch. 2019. \emph{Cochrane
Handbook for Systematic Reviews of Interventions}. John Wiley \& Sons.

\leavevmode\vadjust pre{\hypertarget{ref-horbach2020pandemic}{}}%
Horbach, Serge PJM. 2020. {`Pandemic Publishing: Medical Journals
Strongly Speed up Their Publication Process for COVID-19'}.
\emph{Quantitative Science Studies} 1 (3): 1056--67.

\leavevmode\vadjust pre{\hypertarget{ref-hoy2020rise}{}}%
Hoy, Matthew B. 2020. {`Rise of the Rxivs: How Preprint Servers Are
Changing the Publishing Process'}. \emph{Medical Reference Services
Quarterly} 39 (1): 84--89.

\leavevmode\vadjust pre{\hypertarget{ref-ikonomakis2005text}{}}%
Ikonomakis, M, Sotiris Kotsiantis, and V Tampakas. 2005. {`Text
Classification Using Machine Learning Techniques'}. \emph{WSEAS
Transactions on Computers} 4 (8): 966--74.

\leavevmode\vadjust pre{\hypertarget{ref-james2013introduction}{}}%
James, Gareth, Daniela Witten, Trevor Hastie, and Robert Tibshirani.
2013. \emph{An Introduction to Statistical Learning}. Vol. 112.
Springer.

\leavevmode\vadjust pre{\hypertarget{ref-jayashree2020evaluation}{}}%
Jayashree, P, and PK Srijith. 2020. {`Evaluation of Deep Gaussian
Processes for Text Classification'}. In \emph{Proceedings of the 12th
Language Resources and Evaluation Conference}, 1485--91.

\leavevmode\vadjust pre{\hypertarget{ref-jonnalagadda2015automating}{}}%
Jonnalagadda, Siddhartha R, Pawan Goyal, and Mark D Huffman. 2015.
{`Automating Data Extraction in Systematic Reviews: A Systematic
Review'}. \emph{Systematic Reviews} 4 (1): 1--16.

\leavevmode\vadjust pre{\hypertarget{ref-joo2020being}{}}%
Joo, Taejong, Uijung Chung, and Min-Gwan Seo. 2020. {`Being Bayesian
about Categorical Probability'}. In \emph{International Conference on
Machine Learning}, 4950--61. PMLR.

\leavevmode\vadjust pre{\hypertarget{ref-jospin2020hands}{}}%
Jospin, Laurent Valentin, Wray Buntine, Farid Boussaid, Hamid Laga, and
Mohammed Bennamoun. 2020. {`Hands-on Bayesian Neural Networks--a
Tutorial for Deep Learning Users'}. \emph{arXiv Preprint
arXiv:2007.06823}.

\leavevmode\vadjust pre{\hypertarget{ref-kapelner2013bartmachine}{}}%
Kapelner, Adam, and Justin Bleich. 2013. {`bartMachine: Machine Learning
with Bayesian Additive Regression Trees'}. \emph{arXiv Preprint
arXiv:1312.2171}.

\leavevmode\vadjust pre{\hypertarget{ref-kapelner2015prediction}{}}%
---------. 2015. {`Prediction with Missing Data via Bayesian Additive
Regression Trees'}. \emph{Canadian Journal of Statistics} 43 (2):
224--39.

\leavevmode\vadjust pre{\hypertarget{ref-khoshgoftaar2010comparing}{}}%
Khoshgoftaar, Taghi M, Jason Van Hulse, and Amri Napolitano. 2010.
{`Comparing Boosting and Bagging Techniques with Noisy and Imbalanced
Data'}. \emph{IEEE Transactions on Systems, Man, and Cybernetics-Part A:
Systems and Humans} 41 (3): 552--68.

\leavevmode\vadjust pre{\hypertarget{ref-kohavi1995study}{}}%
Kohavi, Ron et al. 1995. {`A Study of Cross-Validation and Bootstrap for
Accuracy Estimation and Model Selection'}. In \emph{Ijcai}, 14:1137--45.
2. Montreal, Canada.

\leavevmode\vadjust pre{\hypertarget{ref-lai2015recurrent}{}}%
Lai, Siwei, Liheng Xu, Kang Liu, and Jun Zhao. 2015. {`Recurrent
Convolutional Neural Networks for Text Classification'}. In
\emph{Twenty-Ninth AAAI Conference on Artificial Intelligence}.

\leavevmode\vadjust pre{\hypertarget{ref-landauer1998introduction}{}}%
Landauer, Thomas K, Peter W Foltz, and Darrell Laham. 1998. {`An
Introduction to Latent Semantic Analysis'}. \emph{Discourse Processes}
25 (2-3): 259--84.

\leavevmode\vadjust pre{\hypertarget{ref-larsen2010rate}{}}%
Larsen, Peder, and Markus Von Ins. 2010. {`The Rate of Growth in
Scientific Publication and the Decline in Coverage Provided by Science
Citation Index'}. \emph{Scientometrics} 84 (3): 575--603.

\leavevmode\vadjust pre{\hypertarget{ref-laws2008stopping}{}}%
Laws, Florian, and Hinrich Schütze. 2008. {`Stopping Criteria for Active
Learning of Named Entity Recognition'}. In \emph{Proceedings of the 22nd
International Conference on Computational Linguistics (Coling 2008)},
465--72.

\leavevmode\vadjust pre{\hypertarget{ref-lefebvre2011searching}{}}%
Lefebvre, C, E Manheimer, J Glanville, J Higgins, and S Green. 2011.
{`Searching for Studies (Chapter 6)'}. \emph{Cochrane Handbook for
Systematic Reviews of Interventions Version} 510.

\leavevmode\vadjust pre{\hypertarget{ref-li2020survey}{}}%
Li, Qian, Hao Peng, Jianxin Li, Congying Xia, Renyu Yang, Lichao Sun,
Philip S Yu, and Lifang He. 2020. {`A Survey on Text Classification:
From Shallow to Deep Learning'}. \emph{arXiv Preprint arXiv:2008.00364}.

\leavevmode\vadjust pre{\hypertarget{ref-lipscomb2000medical}{}}%
Lipscomb, Carolyn E. 2000. {`Medical Subject Headings (MeSH)'}.
\emph{Bulletin of the Medical Library Association} 88 (3): 265.

\leavevmode\vadjust pre{\hypertarget{ref-malhi2020explainable}{}}%
Malhi, Avleen, Samanta Knapic, and Kary Främling. 2020. {`Explainable
Agents for Less Bias in Human-Agent Decision Making'}. In
\emph{International Workshop on Explainable, Transparent Autonomous
Agents and Multi-Agent Systems}, 129--46. Springer.

\leavevmode\vadjust pre{\hypertarget{ref-marshall2015systematic}{}}%
Marshall, Iain J, and Byron C Wallace. 2019. {`Toward Systematic Review
Automation: A Practical Guide to Using Machine Learning Tools in
Research Synthesis'}. In \emph{Systematic Reviews}, 8:1--10. 1.
Springer.

\leavevmode\vadjust pre{\hypertarget{ref-minaee2021deep}{}}%
Minaee, Shervin, Nal Kalchbrenner, Erik Cambria, Narjes Nikzad, Meysam
Chenaghlu, and Jianfeng Gao. 2021. {`Deep Learning--Based Text
Classification: A Comprehensive Review'}. \emph{ACM Computing Surveys
(CSUR)} 54 (3): 1--40.

\leavevmode\vadjust pre{\hypertarget{ref-miwa2014reducing}{}}%
Miwa, Makoto, James Thomas, Alison O'Mara-Eves, and Sophia Ananiadou.
2014. {`Reducing Systematic Review Workload Through Certainty-Based
Screening'}. \emph{Journal of Biomedical Informatics} 51: 242--53.

\leavevmode\vadjust pre{\hypertarget{ref-moore2018transparent}{}}%
Moore, Alexander, Vanessa Murdock, Yaxiong Cai, and Kristine Jones.
2018. {`Transparent Tree Ensembles'}. In \emph{The 41st International
ACM SIGIR Conference on Research \& Development in Information
Retrieval}, 1241--44.

\leavevmode\vadjust pre{\hypertarget{ref-pubmedUpdate}{}}%
{`NCBI Insights : Updated Pubmed e-Utilities Coming in April 2022!'}.
n.d. U.S. National Library of Medicine.
\url{https://ncbiinsights.ncbi.nlm.nih.gov/2021/10/05/updated-pubmed-api/}.

\leavevmode\vadjust pre{\hypertarget{ref-o2015using}{}}%
O'Mara-Eves, Alison, James Thomas, John McNaught, Makoto Miwa, and
Sophia Ananiadou. 2015. {`Using Text Mining for Study Identification in
Systematic Reviews: A Systematic Review of Current Approaches'}.
\emph{Systematic Reviews} 4 (1): 1--22.

\leavevmode\vadjust pre{\hypertarget{ref-ohsawa1998keygraph}{}}%
Ohsawa, Yukio, Nels E Benson, and Masahiko Yachida. 1998. {`KeyGraph:
Automatic Indexing by Co-Occurrence Graph Based on Building Construction
Metaphor'}. In \emph{Proceedings IEEE International Forum on Research
and Technology Advances in Digital Libraries-ADL'98-}, 12--18. IEEE.

\leavevmode\vadjust pre{\hypertarget{ref-olorisade2016critical}{}}%
Olorisade, Babatunde K, Ed de Quincey, Pearl Brereton, and Peter Andras.
2016. {`A Critical Analysis of Studies That Address the Use of Text
Mining for Citation Screening in Systematic Reviews'}. In
\emph{Proceedings of the 20th International Conference on Evaluation and
Assessment in Software Engineering}, 1--11.

\leavevmode\vadjust pre{\hypertarget{ref-pavlinek2017text}{}}%
Pavlinek, Miha, and Vili Podgorelec. 2017. {`Text Classification Method
Based on Self-Training and LDA Topic Models'}. \emph{Expert Systems with
Applications} 80: 83--93.

\leavevmode\vadjust pre{\hypertarget{ref-poulter2008mscanner}{}}%
Poulter, Graham L, Daniel L Rubin, Russ B Altman, and Cathal Seoighe.
2008. {`MScanner: A Classifier for Retrieving Medline Citations'}.
\emph{BMC Bioinformatics} 9 (1): 1--12.

\leavevmode\vadjust pre{\hypertarget{ref-rstat}{}}%
R Core Team. 2020. \emph{R: A Language and Environment for Statistical
Computing}. Vienna, Austria: R Foundation for Statistical Computing.
\url{https://www.R-project.org/}.

\leavevmode\vadjust pre{\hypertarget{ref-ramezankhani2016impact}{}}%
Ramezankhani, Azra, Omid Pournik, Jamal Shahrabi, Fereidoun Azizi,
Farzad Hadaegh, and Davood Khalili. 2016. {`The Impact of Oversampling
with SMOTE on the Performance of 3 Classifiers in Prediction of Type 2
Diabetes'}. \emph{Medical Decision Making} 36 (1): 137--44.

\leavevmode\vadjust pre{\hypertarget{ref-robert2004monte}{}}%
Robert, Christian P, and George Casella. 2004. \emph{Monte Carlo
Statistical Methods}. Vol. 2. Springer.

\leavevmode\vadjust pre{\hypertarget{ref-roshan2020improvement}{}}%
Roshan, Seyed Ehsan, and Shahrokh Asadi. 2020. {`Improvement of Bagging
Performance for Classification of Imbalanced Datasets Using Evolutionary
Multi-Objective Optimization'}. \emph{Engineering Applications of
Artificial Intelligence} 87: 103319.

\leavevmode\vadjust pre{\hypertarget{ref-rousseau2015graph}{}}%
Rousseau, Francois. 2015. {`Graph-of-Words: Mining and Retrieving Text
with Networks of Features'}. PhD thesis, Ph. D. dissertation.

\leavevmode\vadjust pre{\hypertarget{ref-rousseau2015text}{}}%
Rousseau, François, Emmanouil Kiagias, and Michalis Vazirgiannis. 2015.
{`Text Categorization as a Graph Classification Problem'}. In
\emph{Proceedings of the 53rd Annual Meeting of the Association for
Computational Linguistics and the 7th International Joint Conference on
Natural Language Processing (Volume 1: Long Papers)}, 1702--12.

\leavevmode\vadjust pre{\hypertarget{ref-newis}{}}%
Sadaghiani, Catharina, Tjibbe Donker, Xanthi Andrianou, Balázs Babarczy,
Gerolf De Boer, Francesco Di Ruscio, Shona Cairns, et al. 2020.
\emph{National Health Care Infrastructures, Health Care Utilization and
Patient Movements Between Hospitals - Networks Working to Improve
Surveillance: A Systematic Literature Review}. PROSPERO.
\url{http://www.crd.york.ac.uk/PROSPERO/display_record.asp?ID=CRD42020157987}.

\leavevmode\vadjust pre{\hypertarget{ref-sampson2011precision}{}}%
Sampson, Margaret, Jennifer Tetzlaff, and Christine Urquhart. 2011.
{`Precision of Healthcare Systematic Review Searches in a
Cross-Sectional Sample'}. \emph{Research Synthesis Methods} 2 (2):
119--25.

\leavevmode\vadjust pre{\hypertarget{ref-schonlau2017text}{}}%
Schonlau, Matthias, and Nick Guenther. 2017. {`Text Mining Using
n-Grams'}. \emph{Schonlau, M., Guenther, N. Sucholutsky, I. Text Mining
Using n-Gram Variables. The Stata Journal} 17 (4): 866--81.

\leavevmode\vadjust pre{\hypertarget{ref-settles2009active}{}}%
Settles, Burr. 2009. {`Active Learning Literature Survey'}.

\leavevmode\vadjust pre{\hypertarget{ref-soria2011belm}{}}%
Soria-Olivas, Emilio, Juan Gomez-Sanchis, José D Martin, Joan
Vila-Frances, Marcelino Martinez, José R Magdalena, and Antonio J
Serrano. 2011. {`BELM: Bayesian Extreme Learning Machine'}. \emph{IEEE
Transactions on Neural Networks} 22 (3): 505--9.

\leavevmode\vadjust pre{\hypertarget{ref-soto2019thalia}{}}%
Soto, Axel J, Piotr Przybyła, and Sophia Ananiadou. 2019. {`Thalia:
Semantic Search Engine for Biomedical Abstracts'}. \emph{Bioinformatics}
35 (10): 1799--1801.

\leavevmode\vadjust pre{\hypertarget{ref-tashman2000out}{}}%
Tashman, Leonard J. 2000. {`Out-of-Sample Tests of Forecasting Accuracy:
An Analysis and Review'}. \emph{International Journal of Forecasting} 16
(4): 437--50.

\leavevmode\vadjust pre{\hypertarget{ref-rpart}{}}%
Therneau, Terry, and Beth Atkinson. 2019. \emph{Rpart: Recursive
Partitioning and Regression Trees}.
\url{https://CRAN.R-project.org/package=rpart}.

\leavevmode\vadjust pre{\hypertarget{ref-thomas2007eppi}{}}%
Thomas, James, and Jeff Brunton. 2007. {`EPPI-Reviewer: Software for
Research Synthesis'}.

\leavevmode\vadjust pre{\hypertarget{ref-tsafnat2013automation}{}}%
Tsafnat, Guy, Adam Dunn, Paul Glasziou, and Enrico Coiera. 2013. {`The
Automation of Systematic Reviews'}. British Medical Journal Publishing
Group.

\leavevmode\vadjust pre{\hypertarget{ref-tsafnat2014systematic}{}}%
Tsafnat, Guy, Paul Glasziou, Miew Keen Choong, Adam Dunn, Filippo
Galgani, and Enrico Coiera. 2014. {`Systematic Review Automation
Technologies'}. \emph{Systematic Reviews} 3 (1): 1--15.

\leavevmode\vadjust pre{\hypertarget{ref-turian2010word}{}}%
Turian, Joseph, Lev Ratinov, and Yoshua Bengio. 2010. {`Word
Representations: A Simple and General Method for Semi-Supervised
Learning'}. In \emph{Proceedings of the 48th Annual Meeting of the
Association for Computational Linguistics}, 384--94.

\leavevmode\vadjust pre{\hypertarget{ref-violos2016sentiment}{}}%
Violos, John, Konstantinos Tserpes, Evangelos Psomakelis, Konstantinos
Psychas, and Theodora Varvarigou. 2016. {`Sentiment Analysis Using
Word-Graphs'}. In \emph{Proceedings of the 6th International Conference
on Web Intelligence, Mining and Semantics}, 1--9.

\leavevmode\vadjust pre{\hypertarget{ref-visser2010performing}{}}%
Visser, Eelco. 2010. {`Performing Systematic Literature Reviews with
Researchr: Tool Demonstration'}. \emph{Technical Report Series
TUD-SERG-2010-010}.

\leavevmode\vadjust pre{\hypertarget{ref-wallace2010active}{}}%
Wallace, Byron C, Kevin Small, Carla E Brodley, and Thomas A Trikalinos.
2010. {`Active Learning for Biomedical Citation Screening'}. In
\emph{Proceedings of the 16th ACM SIGKDD International Conference on
Knowledge Discovery and Data Mining}, 173--82.

\leavevmode\vadjust pre{\hypertarget{ref-wallace2010semi}{}}%
Wallace, Byron C, Thomas A Trikalinos, Joseph Lau, Carla Brodley, and
Christopher H Schmid. 2010. {`Semi-Automated Screening of Biomedical
Citations for Systematic Reviews'}. \emph{BMC Bioinformatics} 11 (1):
1--11.

\leavevmode\vadjust pre{\hypertarget{ref-wilkins2005embase}{}}%
Wilkins, Thad, Ralph A Gillies, and Kathy Davies. 2005. {`EMBASE Versus
MEDLINE for Family Medicine Searches: Can MEDLINE Searches Find the
Forest or a Tree?'}. \emph{Canadian Family Physician} 51 (6): 848--49.

\leavevmode\vadjust pre{\hypertarget{ref-woods1998medline}{}}%
Woods, David, and Kate Trewheellar. 1998. {`Medline and Embase
Complement Each Other in Literature Searches'}. \emph{BMJ: British
Medical Journal} 316 (7138): 1166.

\leavevmode\vadjust pre{\hypertarget{ref-yang2020survey}{}}%
Yang, JinXiong, Liang Bai, and Yanming Guo. 2020. {`A Survey of Text
Classification Models'}. In \emph{Proceedings of the 2020 2nd
International Conference on Robotics, Intelligent Control and Artificial
Intelligence}, 327--34.

\leavevmode\vadjust pre{\hypertarget{ref-yang2020hyperparameter}{}}%
Yang, Li, and Abdallah Shami. 2020. {`On Hyperparameter Optimization of
Machine Learning Algorithms: Theory and Practice'}.
\emph{Neurocomputing} 415: 295--316.

\leavevmode\vadjust pre{\hypertarget{ref-zhou2021ensemble}{}}%
Zhou, Zhi-Hua. 2021. {`Ensemble Learning'}. In \emph{Machine Learning},
181--210. Springer.

\leavevmode\vadjust pre{\hypertarget{ref-zhu2010confidence}{}}%
Zhu, Jingbo, Huizhen Wang, Eduard Hovy, and Matthew Ma. 2010.
{`Confidence-Based Stopping Criteria for Active Learning for Data
Annotation'}. \emph{ACM Transactions on Speech and Language Processing
(TSLP)} 6 (3): 1--24.

\end{CSLReferences}

\bibliographystyle{unsrt}
\bibliography{references.bib}

\end{document}

\maketitle

{
\setcounter{tocdepth}{2}
\tableofcontents
}
\newpage

\hypertarget{research-protocol}{%
\section{Research Protocol}\label{research-protocol}}

This document will guide the reader along the steps to utilise the
framework and reproduce our results.

To use the framework clone it from
\url{https://github.com/AD-Papers-Material/BART_SystReviewClassifier}
and run the R project in the repository.

The first step is loading the code base using:

\begin{Shaded}
\begin{Highlighting}[]
\FunctionTok{source}\NormalTok{(}\StringTok{"R/Setup.R"}\NormalTok{) }\CommentTok{\# Load the framework}
\end{Highlighting}
\end{Shaded}

\hypertarget{acquisition-of-the-citation-data}{%
\subsection{Acquisition of the citation
data}\label{acquisition-of-the-citation-data}}

Once the framework is loaded, the user defines an initial search query
which needs to be as specific as possible while still generating a
sufficient number of positive matches:

\begin{Shaded}
\begin{Highlighting}[]
\CommentTok{\# Initial query to be built on domain knowledge. It accepts OR, AND, NOT boolean}
\CommentTok{\# operators and round brackets to group terms.}
\NormalTok{query }\OtherTok{\textless{}{-}} \StringTok{\textquotesingle{}((model OR models OR modeling OR network OR networks) AND}
\StringTok{(dissemination OR transmission OR spread OR diffusion) AND (nosocomial OR}
\StringTok{hospital OR "long{-}term{-}care" OR "long term care" OR "longterm care" OR}
\StringTok{"long{-}term care" OR "healthcare associated") AND (infection OR resistance OR}
\StringTok{resistant))\textquotesingle{}}

\CommentTok{\# Year filter. The framework converts it to the API{-}specific format seamlessly.}
\CommentTok{\# common logical comparators can be used, i.e. \textless{}, \textless{}=, \textgreater{}, \textgreater{}=, while dashes}
\CommentTok{\# denotes inclusive date intervals. A single year restricts results to one year}
\CommentTok{\# period.}
\NormalTok{year\_filter }\OtherTok{\textless{}{-}} \StringTok{"2010{-}2020"}
\end{Highlighting}
\end{Shaded}

The query is passed to the \texttt{perform\_search\_session()} function,
together with an unique identifier for the \textbf{search session} and
\textbf{search query}. A session identifies a homogeneous group of
complementary or alternative queries (e.g.~sometimes it is easier to
create two different simpler queries than a complex and long one that
returns the same results).\\
In our implementation, we define as sessions the group of actions that
includes the utilisation of a search query and the subsequent iterations
of manual review and automatic classification of its results.\\

\texttt{perform\_search\_session()} will use common scientific research
database APIs (at the moment Pubmed/Medline, Web of Science and IEEE) to
acquire records related to the query. The records will be stored as CSV
files in folders with the following structure:
\texttt{Records/session\_id/query\_id}, created automatically. The
function also creates or updates a \textbf{journal} (a CSV or Microsoft
Excel file) with all the information about sessions, queries and search
results.\\
Some services (IEEE, WOS) requires an API key to access the data; the
API key is not mandatory for Pubmed but can help avoid quota limitations
in case of a large number of results. IEEE can also be searched without
API: the framework will utilise a web scraping approach if the API key
is missing, but the success of this method is not guaranteed. See
Supplemental Material S3 for detailed instruction on preliminary steps
and tips for each supported search engine. These keys can be stored in
the \texttt{secrets.R} file in the working directory, with the following
example code (the ``\emph{baysren}'' prefix identifies the package
specific options):

\begin{Shaded}
\begin{Highlighting}[]
\FunctionTok{options}\NormalTok{(}\AttributeTok{baysren.ieee\_api\_key =} \StringTok{"your{-}ieee{-}api{-}key"}\NormalTok{)}
\FunctionTok{options}\NormalTok{(}\AttributeTok{baysren.ncbi\_api\_key =} \StringTok{"your{-}ncbi{-}api{-}key"}\NormalTok{)}
\FunctionTok{options}\NormalTok{(}\AttributeTok{baysren.wos\_api\_key =} \StringTok{"your{-}wos{-}api{-}key"}\NormalTok{)}
\end{Highlighting}
\end{Shaded}

This file will be automatically parsed from the framework.

If previously downloaded results are already available, users can create
the \texttt{Record} folder manually and put the data there (the file
names need to contain the source name as written in the \texttt{sources}
argument of \texttt{perform\_search\_session()}). The function will
acquire and parse them into a standard format, even if split into
multiple files due to download limits (e.g.~\texttt{Pubmed1.nbib},
\texttt{Pubmed2.nib}, etc.). In addition to Pubmed, WOS and IEEE, this
method permits to import also EMBASE and SCOPUS records, for which an
API search was not yet implemented.\\
Using the \texttt{actions} argument, users can choose whether to perform
an API search, parse already downloaded results, or both.

Note that for Pubmed, the API search may return slightly fewer records
than a manual one due to a different query expansion algorithm (this
divergence should be solved in April 2022) between the two services,
therefore is advisable to perform also a manual search and put the
resulting .nbib files in the \texttt{Session/Query} folder.
\texttt{perform\_search\_session()} will acquire them seamlessly.

If a user wants to perform the API searches and results parsing manually
for each source, the framework exposes the individual functions to
search for records using the APIs and parsing bibliography files.

\begin{Shaded}
\begin{Highlighting}[]
\CommentTok{\# This call will perform an API search on Pubmed, WOS and IEEE services and/or}
\CommentTok{\# (based on the \textasciigrave{}actions\textasciigrave{} argument) parse already downloaded data (in our case,}
\CommentTok{\# additional Pubmed results and Embase and Scopus results) manually added to the}
\CommentTok{\# Session1/Query1 folder.}

\NormalTok{journal }\OtherTok{\textless{}{-}} \FunctionTok{perform\_search\_session}\NormalTok{(}
  \AttributeTok{query =}\NormalTok{ query, }\AttributeTok{year\_query =}\NormalTok{ year\_filter,}
  \AttributeTok{session\_name =} \StringTok{"Session1"}\NormalTok{, }\AttributeTok{query\_name =} \StringTok{"Query1"}\NormalTok{,}
  \AttributeTok{records\_folder =} \StringTok{"Records"}\NormalTok{,}
  \AttributeTok{journal =} \StringTok{"Session\_journal.csv"}
\NormalTok{)}
\end{Highlighting}
\end{Shaded}

Once the records are stored, they must be read and merged into an
\textbf{annotation file}, where the initial manual review will be
performed.\\
A series of functions are available to prepare the data for manual
evaluation:

\begin{Shaded}
\begin{Highlighting}[]
\CommentTok{\# Extract the file paths of records. Arguments can be used to filter by session}
\CommentTok{\# query, source. Only parsed files will be returned, not the raw ones downloaded}
\CommentTok{\# manually.}
\NormalTok{Annotation\_data }\OtherTok{\textless{}{-}} \FunctionTok{extract\_source\_file\_paths}\NormalTok{(journal) }\SpecialCharTok{\%\textgreater{}\%}
  \CommentTok{\# Read record files. Will parse them if raw data is downloaded manually (not}
  \CommentTok{\# necessary in this case). Return a list of records, one per file.}
  \FunctionTok{read\_bib\_files}\NormalTok{() }\SpecialCharTok{\%\textgreater{}\%}
  \CommentTok{\# Join a list of records into one dataframe, solving duplicated records}
  \FunctionTok{join\_records}\NormalTok{() }\SpecialCharTok{\%\textgreater{}\%}
  \CommentTok{\# Order records by the relative frequency of the query terms in the title +}
  \CommentTok{\# abstract text. Increases the chance of encountering relevant records at the}
  \CommentTok{\# beginning of the manual classification.}
  \FunctionTok{order\_by\_query\_match}\NormalTok{(query) }\SpecialCharTok{\%\textgreater{}\%}
  \CommentTok{\# Add fields for the manual classification}
  \FunctionTok{mutate}\NormalTok{(}
    \AttributeTok{Rev\_manual =} \ConstantTok{NA}\NormalTok{, }\CommentTok{\# Where the first manual classification will be made}
    \AttributeTok{Rev\_prediction =} \ConstantTok{NA}\NormalTok{, }\CommentTok{\# Column for the evaluation of the predicted classes}
    \AttributeTok{.before =}\NormalTok{ DOI}
\NormalTok{  )}
\end{Highlighting}
\end{Shaded}

All these steps are joined into the \texttt{create\_annotation\_file()}
function. This function can take as input either the path to record
files, or the parsed records (one data frame or a list of data frames)
or a group of folders where to look for files into./ In addition, it
also allows adding the new records to a previous annotation file (arg
\texttt{prev\_records}). This is useful when performing consecutive
search sessions with different queries; previous manual classifications
can be imported using \texttt{prev\_classification}.\\
If the original query is passed to the \texttt{reorder\_query} argument,
the records are reordered according to the frequency of query terms in
the title and abstract (\textbf{simple query ordering}); this approach
lets increase the rate of relevant matches at the beginning of the
dataset.

\begin{Shaded}
\begin{Highlighting}[]
\CommentTok{\# This function extracts the appropriate file paths from a session journal. By}
\CommentTok{\# default it passes all stored files, otherwise it can be filtered by session,}
\CommentTok{\# query, and source (i.e. Pubmed, WOS, IEEE)}
\NormalTok{record\_files }\OtherTok{\textless{}{-}} \FunctionTok{extract\_source\_file\_paths}\NormalTok{(journal)}

\CommentTok{\# create\_annotation\_file() accept a great variety of inputs:}

\CommentTok{\# either record file paths}
\NormalTok{input }\OtherTok{\textless{}{-}}\NormalTok{ record\_files}

\CommentTok{\# or specific record file folders}
\NormalTok{input }\OtherTok{\textless{}{-}} \FunctionTok{file.path}\NormalTok{(}\StringTok{"Records"}\NormalTok{, }\StringTok{"Session1"}\NormalTok{, }\StringTok{"Query1"}\NormalTok{)}

\CommentTok{\# or parent folders, since it searches for files recursively}
\NormalTok{input }\OtherTok{\textless{}{-}} \StringTok{"Records"}

\CommentTok{\# or the already parsed files}
\NormalTok{input }\OtherTok{\textless{}{-}} \FunctionTok{read\_bib\_files}\NormalTok{(record\_files)}

\CommentTok{\# We can then call create\_annotation\_file() with one of the above input}
\NormalTok{Annotation\_data }\OtherTok{\textless{}{-}} \FunctionTok{create\_annotation\_file}\NormalTok{(input, }\AttributeTok{reorder\_query =}\NormalTok{ query)}
\end{Highlighting}
\end{Shaded}

Once created, the annotations need to be stored in an \textbf{Annotation
Session} folder, which will contain all information necessary for the
automatic classification. This operation is streamlined by the
\texttt{create\_session()} function, which also helps manage duplicated
sessions.

\begin{Shaded}
\begin{Highlighting}[]
\CommentTok{\# Create the first session}
\FunctionTok{create\_session}\NormalTok{(Annotation\_data, }\AttributeTok{session\_name =} \StringTok{"Session1"}\NormalTok{)}
\end{Highlighting}
\end{Shaded}

The file annotation file is called \texttt{Records\_\{date\}}, situated
in the folder \texttt{Sessions/\{session\_name\}}; \texttt{date} here
represents the session creation timestamp, and \texttt{session\_name}
the name passed by the user.

Notice that the framework is ``pipe'' friendly, and the precedent steps
can be all performed in one call:

\begin{Shaded}
\begin{Highlighting}[]
\CommentTok{\# First, create the Records/Session1/Query1 folder and put the manually}
\CommentTok{\# downloaded records there if there are any. Otherwise, the folders will be}
\CommentTok{\# created automatically during the API search}

\CommentTok{\# perform searches and parse files; save and return the journal}
\NormalTok{Records }\OtherTok{\textless{}{-}} \FunctionTok{perform\_search\_session}\NormalTok{(}
  \AttributeTok{query =}\NormalTok{ query, }\AttributeTok{year\_query =}\NormalTok{ year\_filter,}
  \AttributeTok{session\_name =} \StringTok{"Session1"}\NormalTok{, }\AttributeTok{query\_name =} \StringTok{"Query1"}\NormalTok{,}
  \AttributeTok{journal =} \StringTok{"Session\_journal.csv"}
\NormalTok{) }\SpecialCharTok{\%\textgreater{}\%}
  \CommentTok{\# return the record file paths}
  \FunctionTok{extract\_source\_file\_paths}\NormalTok{() }\SpecialCharTok{\%\textgreater{}\%}
  \CommentTok{\# read, join, reorder records and save them in the Sessions folder}
  \FunctionTok{create\_annotation\_file}\NormalTok{(}\AttributeTok{reorder\_query =}\NormalTok{ query) }\SpecialCharTok{\%\textgreater{}\%}
  \CommentTok{\# Create a new session with the records}
  \FunctionTok{create\_session}\NormalTok{(}\AttributeTok{session\_name =} \StringTok{"Session1"}\NormalTok{)}
\end{Highlighting}
\end{Shaded}

\hypertarget{manual-classification-and-prediction}{%
\subsection{Manual classification and
prediction}\label{manual-classification-and-prediction}}

Now the annotation file is ready for the manual classification of
records. The annotation file has a column called \texttt{Rev\_manual},
where records need to be classified as ``positive'' (y) or ``negative''
(n) matches by the user. We suggest manually labelling an
\textbf{initial training set} of 250 records, which should be sufficient
if a minimum \textasciitilde20\% positivity rate is present (see the
section about model hyperparameter optimisation)./ The function
\texttt{plot\_classification\_trends()} can help visualise the trend of
the positives/negatives as the number of classified records increases./

Once the initial training set is prepared, the function
\texttt{enrich\_annotation\_file()} will use a Bayesian Additive
Regression Trees (BART) machine learning model to learn the labelling
pattern given the record data and predict the positive match probability
for the unlabelled records.\\
\texttt{enrich\_annotation\_file()} comprises two steps. The first step
is the creation of a \textbf{Document Term Matrix} (DTM):

\begin{itemize}
\tightlist
\item
  The framework extracts \textbf{features} from text data (i.e., title
  and abstract) of the records, authors, keywords and MESH terms (see
  Methods for more info). Before features extraction, text data terms
  are lemmatised, stop-words are removed, and only nouns, adjectives,
  verbs and untagged terms are retained. Authors, keywords, and MESH are
  kept unmodified. Co-occurrent features (non-consecutive ngrams,
  \textbf{nc-grams}) are identified and \textbf{redundant features}
  (i.e.~terms always appearing together) are joined to decrease the
  feature space noise.
\item
  For each term, a 0/1 label specifies if it is present in each record.
\item
  Positive records are oversampled \textbf{10x} to improve sensitivity;
\item
  An extra feature is added for each record field (title, abstract,
  authors, keywords, MESH terms), reporting the number of terms in each.
\end{itemize}

The second step is the modelling of the labelling pattern and the
predictions of the relevancy/positivity probability:

\begin{itemize}
\tightlist
\item
  The function models the labelling pattern using the manually labelled
  records (**learning phaseéé);
\item
  After the model is created, it assigns a \textbf{posterior predictive
  distribution} (PPD) of the probability of positivity to each record
  (\textbf{prediction phase}). If the number of records is large, the
  prediction step is broken into chunks of \textbf{5000 records}, to
  avoid memory saturation; the users can set up this parameter according
  to their machine capabilities;
\item
  This process is repeated \textbf{10 times}, and the predictive
  distributions are averaged to induce shrinkage of the estimates (i.e.,
  decrease the chance of outliers), creating a prediction
  \textbf{ensemble};
\item
  The resulting distributions are summarised by the \textbf{98\%
  interval} {[}98\% PrI{]} and the median; the {[}98\% PrI{]} identifies
  the \textbf{uncertainty zone};
\item
  The unlabelled records will receive a positive/negative predicted
  label; the label is positive if its lower {[}98\% PrI{]} bound is
  higher than the uncertainty zone upper bound and negative if the upper
  {[}98\% PrI{]} bound is lower than the uncertainty zone lower bound.
  All other records are labelled as \textbf{uncertain} (unk);
\item
  If a predicted label does not match an existing one, the record will
  be labelled as \textbf{check}.
\end{itemize}

The user can define parameters like the positives' oversampling rate,
the number of models to average in the ensemble, whether to apply
bootstrap resampling before each repetition, the prediction chunk size,
and the PPD quantiles used to build the uncertainty zone. To use the
function, the only input needed is the session name; the function will
automatically identify the relevant file to make predictions on.

\begin{Shaded}
\begin{Highlighting}[]
\NormalTok{New\_annotations }\OtherTok{\textless{}{-}} \FunctionTok{enrich\_annotation\_file}\NormalTok{(}\StringTok{"Session1"}\NormalTok{)}
\end{Highlighting}
\end{Shaded}

This function will produce a new file which is stored in the Annotations
folder of the session; the file will contain a column called
\texttt{Rev\_prediction\_new} where the user will find records that
require manual review marked by ``*``. These are records with uncertain
labelling (''unk''), but also positive predictions and ``y''/``n''
predictions that contradict previous manually input labels (labelled as
``check''). This approach increases the accuracy of the predictions
since the machine learning system is set up to favour sensitivity at the
expense of specificity. If the function finds prelabelled records
(\texttt{Rev\_previous} column) added when the Annotation file was
created or passed as a further input to
\texttt{enrich\_annotation\_file()} (\texttt{prev\_classification}
argument), they will be used to label the records marked in the
\texttt{Rev\_prediction\_new} column.

The function adds a number of diagnostic pages to the annotation file:

\begin{itemize}
\tightlist
\item
  A results summary page reporting the number of records in need of
  manual review (which includes ``unk'', ``check'', and unreviewed ``y''
  records), the changes, and the final distribution of the labels. It
  also shows other information like the number of features in the DTM,
  the path to the parent file from which the annotation file was
  created, the replication number (see later). The content of this
  summary is also saved as a CSV file in the Results folder.
\item
  A variable importance page that lists the used terms in order of the
  fraction of BART trees in which they were used. If an ensemble of
  models is used, a more robust score is computed, taking the ratio of
  the mean of the term tree fraction among models and its standard
  deviation.
\item
  The arguments passed to the function if different from the default,
  useful for reproducibility.
\item
  A classification performance summary. The procedure may take some time
  and requires \texttt{cmdstan} installed on the machine. Therefore the
  \texttt{compute\_performance} argument is set to \texttt{FALSE} by
  default.
\end{itemize}

Finally, the PPD ensemble produced by the Bayesian engine is stored on
the disk. This feature can be turned off to save disk space, but at
least the samples of the last annotation are necessary to build a new
search query (see next section).

The process of predicting labels and manually reviewing them is what
defines a ``classification-review (CR) iteration''.

Once the manual review of the generated annotation file is completed,
the user can \textbf{rerun exactly the same function}. The last reviewed
annotation file will be automatically picked up and used for a new CR
iteration. If the number of records requiring manual review is
excessively high (e.g.~\textgreater{} 250), the function needs to be
called with the \texttt{stop\_on\_unreviewed} argument set to
\texttt{FALSE} to allow the classification to proceed even in the
presence of unreviewed records in the \texttt{Rev\_prediction\_new}
column.

If there are no positive results (after manual review) in a CR
iteration, the framework records the next iteration as a
\textbf{replication}. After four replications with no new positive
matches, the function will prevent another iteration and send a warning
message, indicating that the session is complete. The user can change
the replication limit; furthermore, it is possible to prevent the
function from starting if either a specified number (or a fraction) of
the total records has been manually reviewed or if a certain number of
positive matches have been found. These three parameters are passed to
the \texttt{limits} argument, which has the following structure and
defaults:

\begin{Shaded}
\begin{Highlighting}[]
\FunctionTok{list}\NormalTok{(}
  \AttributeTok{stop\_after =} \DecValTok{4}\NormalTok{,}
  \AttributeTok{pos\_target =} \ConstantTok{NULL}\NormalTok{,}
  \AttributeTok{labeling\_limit =} \ConstantTok{NULL}
\NormalTok{)}
\end{Highlighting}
\end{Shaded}

The result files in the Results folder are generated after the
Classification step but before the Review one, and therefore show
preliminary results with still unreviewed records. The function
\texttt{consolidate\_results(session\_name)} will update the results
files to show the results after the manual review. The original result
data is stored anyway in the respective annotations files.

\hypertarget{new-search-query-generation}{%
\subsection{New search query
generation}\label{new-search-query-generation}}

It is not uncommon for the first query used in a systematic review not
to be the most effective one, especially in terms of sensitivity. We
tried to address the problem by creating an automatic query generation
system.\\
The solution is experimental and greatly favours sensitivity at the
expense of specificity. The general query generation mechanism is the
following:

\begin{itemize}
\tightlist
\item
  We fit a partition tree between the DTM and 800 samples from the PPD;
  if a term is present multiple times in the DTM (e.g., both title and
  abstract), they are counted just one, and field term count features
  are removed. This step generates a list of rules composed by
  \(AND\)/\(NOT\) ``conditions'' made of terms/authors/keywords/MESH
  tokens, which together identify a group of records.
\item
  For each rule, negative conditions (i.e., \(NOT\) statements) are
  added iteratively, starting from the most specific one, until no
  conditions are found that would not also remove positive records.
\item
  The extended set of rules is sorted by positive-negative record
  difference in descending order. The cumulative number of unique
  positive records is computed and used to group the rules. Rules inside
  each group are ordered by specificity.
\item
  The researcher is then asked to review the rule groups, selecting one
  or more rules (useful if they convey different meaning) from each, or
  edit them (in case too specific positive or negative conditions were
  included). It is possible to exclude a group of rules altogether,
  especially those with the worse sensitivity/specificity ratio.
\item
  The selected rules are joined together by \(OR\) statements, defining
  a subset of records with a sensibly higher proportion of positive
  records than the original one.
\item
  Redundant rules (i.e., rules whose positive records are already
  included in more specific ones) and conditions (i.e., conditions that
  once removed do not decrease the total number of positive or do not
  increase the negative records) are removed.
\item
  Finally, the rules are re-elaborated in a query usable on the major
  scientific databases.
\end{itemize}

The steps for generating a query are the following:

\begin{Shaded}
\begin{Highlighting}[]
\CommentTok{\# The last annotation file and posterior sample matrix in the session are found;}
\CommentTok{\# the most predictive terms (based on the importance score) are chosen and}
\CommentTok{\# simple prediction trees are used to map the terms to a random subset of the}
\CommentTok{\# posterior probabilities (extracted from the posterior sample matrix saved}
\CommentTok{\# during the last CR iteration). The score threshold and the number of trees can}
\CommentTok{\# be chosen by the user; less stringent values can significantly increase}
\CommentTok{\# computation time. By default results will be saved in a file called}
\CommentTok{\# rule\_data.rds in the session folder.}

\NormalTok{candidate\_queries }\OtherTok{\textless{}{-}} \FunctionTok{extract\_rules}\NormalTok{(}\StringTok{"Session1"}\NormalTok{)}

\CommentTok{\# The \textquotesingle{}candidate\_queries\textquotesingle{} contains the rules, the DMT and the labels on which}
\CommentTok{\# the rules were built, which must be passed to generate\_rule\_selection\_set().}
\CommentTok{\# This function selects the most effective (higher positive/negative matches}
\CommentTok{\# delta) rules; also adds negative terms to increase specificity. A data frame}
\CommentTok{\# is generated (or automatically saved if save\_path is set). The output proposes}
\CommentTok{\# groups of equivalent (in terms of the number of positive matches found) groups}
\CommentTok{\# of query, ordered by specificity; for each group, the user needs to select the}
\CommentTok{\# best rule by setting the selected\_rule column to TRUE. The groups are in order}
\CommentTok{\# of growing sensitivity/decreasing specificity, so it may make sense to exclude}
\CommentTok{\# the last groups totally in order to avoid too many search results.}

\NormalTok{Target }\OtherTok{\textless{}{-}}\NormalTok{ candidate\_queries}\SpecialCharTok{$}\NormalTok{DTM}\SpecialCharTok{$}\NormalTok{Target}
\NormalTok{SpecificDTM }\OtherTok{\textless{}{-}}\NormalTok{ candidate\_queries}\SpecialCharTok{$}\NormalTok{SpecificDTM}

\CommentTok{\# No default is given for the save file, since generate\_rule\_selection\_set()}
\CommentTok{\# have no info about the session it\textquotesingle{}s working on.}
\NormalTok{selection\_set\_file }\OtherTok{\textless{}{-}} \FunctionTok{file.path}\NormalTok{(}\StringTok{"Sessions"}\NormalTok{, }\StringTok{"Session1"}\NormalTok{, }\StringTok{"Selected\_rules.xlsx"}\NormalTok{)}

\NormalTok{selection\_set }\OtherTok{\textless{}{-}} \FunctionTok{generate\_rule\_selection\_set}\NormalTok{(}
\NormalTok{  candidate\_queries}\SpecialCharTok{$}\NormalTok{rule,}
  \AttributeTok{target\_vec =}\NormalTok{ Target,}
  \AttributeTok{target\_data =}\NormalTok{ SpecificDTM,}
  \AttributeTok{save\_path =}\NormalTok{ selection\_set\_file}
\NormalTok{)}

\CommentTok{\# Once the manual selection is performed the selected rules can be further}
\CommentTok{\# simplified in order to remove redundant rules and terms. The aim is to provide}
\CommentTok{\# the shortest rule without loss of sensitivity.}

\NormalTok{simplified\_rules }\OtherTok{\textless{}{-}} \FunctionTok{file.path}\NormalTok{(}\StringTok{"Sessions"}\NormalTok{, }\StringTok{"Session1"}\NormalTok{, }\StringTok{"Selected\_rules.xlsx"}\NormalTok{) }\SpecialCharTok{\%\textgreater{}\%}
  \FunctionTok{import\_data}\NormalTok{() }\SpecialCharTok{\%\textgreater{}\%}
  \FunctionTok{simplify\_ruleset}\NormalTok{(}\AttributeTok{target\_vec =}\NormalTok{ Target, }\AttributeTok{target\_data =}\NormalTok{ SpecificDTM) }\SpecialCharTok{\%\textgreater{}\%}
  \FunctionTok{pull}\NormalTok{(rule)}

\CommentTok{\# Finally, the selected rules need to be joined in a query that scientific}
\CommentTok{\# search engines can understand}

\NormalTok{new\_query }\OtherTok{\textless{}{-}} \FunctionTok{rules\_to\_query}\NormalTok{(simplified\_rules)}
\end{Highlighting}
\end{Shaded}

One last thing to keep in mind is that the query generation algorithm is
only aware of the literature collected inside the annotation file. So it
will only be able to pick term combinations that identify positive
matches inside the already specific set of documents while not
necessarily being discriminant from the rest of the global literature.\\
Once the final query is generated, it is advisable to add a constraining
subquery with an AND logic to restrict the results to the scope of the
topic of interest.\\
In our case, we added ``AND ((antimicrobial resistance) OR (healthcare
infection))'' since, being the global topic of our research, those terms
were understandably not selected by the query generation algorithm as
discriminant.\\

The final query was the following:

\begin{Shaded}
\begin{Highlighting}[]
\NormalTok{query }\OtherTok{\textless{}{-}} \StringTok{"(((Donker T) NOT (bacterium isolate)) OR ((network patient) AND}
\StringTok{(resistant staphylococcus aureus) NOT (monte carlo) NOT isolation) OR}
\StringTok{(facility AND (network patient) AND regional NOT hospitals NOT increase NOT}
\StringTok{(patient transport) NOT (control infection use)) OR ((patient transport) NOT}
\StringTok{(Donker T) NOT worker) OR (hospitals AND (network patient) NOT}
\StringTok{(patient transport) NOT regional NOT clinical) OR (facility AND}
\StringTok{(network patient) NOT hospitals NOT (patient transport) NOT regional NOT}
\StringTok{prevention NOT medical) OR ((healthcare facility) NOT (Donker T) NOT}
\StringTok{worker NOT positive) OR (hospitals NOT (network patient) NOT medical NOT}
\StringTok{environmental NOT outcome NOT global) OR ((network patient) NOT facility NOT}
\StringTok{hospitals NOT (patient transport) NOT therapy NOT global)) AND}
\StringTok{((antimicrobial resistance) OR (healthcare infection))"}
\end{Highlighting}
\end{Shaded}

Another advantage of this technique is an at-glance view of relevant
concepts for the topic of interest.

\hypertarget{new-search-session}{%
\subsection{New search session}\label{new-search-session}}

Once the new query was generated, a second search session can be
created, using the strategy described in the protocol. Due to the lower
specificity of the second query, particular care may be needed in
managing the far larger amount of records: multiple record files may
need to be downloaded from Pubmed, EMBASE, and SCOPUS since they have
limits on the number of records one can download at once; also, the
machine learning classification processing time increases proportionally
with the number of records to label.\\
It is advisable to consider removing the less specific rules iteratively
from the new query until a manageable number of records is obtained,
considering that the probability of finding new positive matches not
already found by more specific queries drops exponentially.\\

Once the new records are collected are manually collected for databases
without an API access, a new search can be performed. The following code
describe how to create a new session. Keep in mind to give the session a
new name and to import the annotated records from the last session when
generating the new annotation file.

\begin{Shaded}
\begin{Highlighting}[]
\CommentTok{\# Repeat the search with the new query, giving a different Session name not to}
\CommentTok{\# override previous records}
\NormalTok{journal }\OtherTok{\textless{}{-}} \FunctionTok{perform\_search\_session}\NormalTok{(}
  \AttributeTok{query =}\NormalTok{ query, }\AttributeTok{year\_query =}\NormalTok{ year\_filter,}
  \AttributeTok{session\_name =} \StringTok{"Session2"}\NormalTok{, }\AttributeTok{query\_name =} \StringTok{"Query1"}\NormalTok{,}
  \AttributeTok{records\_folder =} \StringTok{"Records"}\NormalTok{,}
  \AttributeTok{journal =} \StringTok{"Session\_journal.csv"}
\NormalTok{)}

\CommentTok{\# Get the path to the annotated data in the previous session}
\NormalTok{previousAnnotations }\OtherTok{\textless{}{-}} \FunctionTok{get\_session\_files}\NormalTok{(}\StringTok{"Session1"}\NormalTok{)}\SpecialCharTok{$}\NormalTok{Annotations }\SpecialCharTok{\%\textgreater{}\%} \FunctionTok{last}\NormalTok{()}

\CommentTok{\# Create the new annotation file passing the folder with the new records}
\NormalTok{Annotation\_data }\OtherTok{\textless{}{-}} \FunctionTok{create\_annotation\_file}\NormalTok{(}
  \FunctionTok{file.path}\NormalTok{(}\StringTok{"Records"}\NormalTok{, }\StringTok{"Session2"}\NormalTok{),}
  \AttributeTok{reorder\_query =}\NormalTok{ query,}
  \AttributeTok{prev\_records =}\NormalTok{ previousAnnotations}
\NormalTok{)}

\CommentTok{\# Create the new annotation session}
\FunctionTok{create\_session}\NormalTok{(Annotation\_data, }\AttributeTok{session\_name =} \StringTok{"Session2"}\NormalTok{)}
\end{Highlighting}
\end{Shaded}

Once the new annotation session data is created,
\texttt{enrich\_annotation\_file(\textquotesingle{}Session2\textquotesingle{})}
can be used to perform the CR iterations. It may be possible that during
the first CR iteration, the number of records requiring manual review is
very large: we suggest evaluating still not more than 250, setting the
\texttt{stop\_on\_unreviewed} argument to \texttt{FALSE} to proceed in
the automatic classification even if some uncertain records from the
precedent iteration are still unreviewed.

\hypertarget{result-analysis-and-reporting}{%
\subsection{Result analysis and
reporting}\label{result-analysis-and-reporting}}

Several tools are available to evaluate the classification
performance.\\
As said before, a result file is stored in the session folder for each
CR iteration with information on the classification status; remember to
run \texttt{consolidate\_results(session\_name)} if you want these
results to account for the manual review of the predicted labels.\\
\texttt{compute\_changes()} is a simple function that take an annotation
data frame and computes the changes from the last annotation
iteration.\\
The primary function for result analysis is
\texttt{estimate\_performance()}. The function uses a Bayesian model to
estimate how the probability of a positive match drops as records' lower
boundary of the posterior predictive interval decreases. This model
computes the cumulative distribution of possibly missed positive
records, which permits extraction of the engine's theoretical
sensitivity and efficiency to lower 90\% credibility boundaries (i.e.,
the indicators are better than these values with 95\% probability).
These boundaries depict worse case scenarios since it is impossible to
know the actual number of missed positive records.\\
The function returns the Bayesian \(R^2\) of the model (values below
85-90\% would invalidate the model analysis), the posterior sensitivity,
the observed and theoretical number positives, the expected efficiency
given the theoretical number of positives (that is 1 - the number of
reviewed records above the number of records to review to find the
predicted number of positives according to a negative hypergeometrical
distribution). Also, the function plots the cumulative distribution of
the observed and theoretical number of positive records.\\
\texttt{extract\_var\_imp()} extract the most relevant terms used by the
BART model and summarises their relevancy for the machine learning model
and their statistical relevancy in a simple generalised linear model. A
discrepancy between these two scores indicates a non-linear
effect/interaction between terms.

Finally, several functions are available (check \texttt{R/Reporting.R})
to summarise and format the results.\\
For example, \texttt{summarise\_by\_source()} and
\texttt{summarise\_sources\_by\_session()}, respectively for one session
or a group of sessions, report the scientific databases the records are
collected from, while \texttt{get\_source\_distribution()} describes the
distribution of the number of common sources per record.\\
\texttt{summarise\_annotations()} and
\texttt{summarise\_annotations\_by\_session()} describe the results of
the CR iterations, again for one or multiple sessions, reporting the
number of positive and negative matches, the number of terms (features)
used by the machine learning model, and how the labels change between
iterations. \texttt{format\_performance()} and
\texttt{format\_var\_imp()} format the output of the similarly named
analyses. \texttt{plot\_predictive\_densities()} plots how the mixture
of posterior predictive distributions of records, divided into positive,
negative and to-review records, change between iterations.

\hypertarget{model-parameters-optimisation}{%
\section{Model parameters'
optimisation}\label{model-parameters-optimisation}}

As with many complex machine learning models, the choice of
hyperparameters can significantly influence the framework performance.
For a systematic review helper algorithm, we define the performance as
the ability to find as many positive matches by manually
labelling/reviewing as few records as possible:\\

\[
Sensitivity \times Efficiency =
\frac{Pos.\ found}{Tot.\ pos.} \times \bigg(1 - \frac{Records\ reviewed}{Tot.\ records}\bigg)
\]

We utilise a ``grid search'' algorithm to find the best set of
hyperparameters. We tested these tools on our dataset, manually
labelling the first 1200 records, ordered by simple query match.\\
The first step is to create a folder to store the file required for the
search, for example, \emph{Grid Search}. Then a fully reviewed record
file in the format generated by \texttt{create\_annotation\_file()} is
necessary; this file can be put in the \emph{Grid Search}, but it is not
mandatory.\\

Once the folder and the file are ready, run:

\begin{Shaded}
\begin{Highlighting}[]
\NormalTok{Grid\_search }\OtherTok{\textless{}{-}} \FunctionTok{perform\_grid\_evaluation}\NormalTok{(}
\NormalTok{  records,}
  \AttributeTok{sessions\_folder =} \StringTok{"Grid\_Search"}\NormalTok{,}
  \AttributeTok{prev\_classification =}\NormalTok{ records,}
  \DocumentationTok{\#\# Model parameters (can be changed by users)}
  \AttributeTok{resample =} \FunctionTok{c}\NormalTok{(}\ConstantTok{FALSE}\NormalTok{, }\ConstantTok{TRUE}\NormalTok{),}
  \AttributeTok{n\_init =} \FunctionTok{c}\NormalTok{(}\DecValTok{50}\NormalTok{, }\DecValTok{100}\NormalTok{, }\DecValTok{250}\NormalTok{, }\DecValTok{500}\NormalTok{),}
  \AttributeTok{n\_models =} \FunctionTok{c}\NormalTok{(}\DecValTok{1}\NormalTok{, }\DecValTok{5}\NormalTok{, }\DecValTok{10}\NormalTok{, }\DecValTok{20}\NormalTok{, }\DecValTok{40}\NormalTok{, }\DecValTok{60}\NormalTok{),}
  \AttributeTok{pos\_mult =} \FunctionTok{c}\NormalTok{(}\DecValTok{1}\NormalTok{, }\DecValTok{10}\NormalTok{, }\DecValTok{20}\NormalTok{),}
  \AttributeTok{pred\_quants =} \FunctionTok{list}\NormalTok{(}
    \FunctionTok{c}\NormalTok{(.}\DecValTok{1}\NormalTok{, .}\DecValTok{5}\NormalTok{, .}\DecValTok{9}\NormalTok{),}
    \FunctionTok{c}\NormalTok{(.}\DecValTok{05}\NormalTok{, .}\DecValTok{5}\NormalTok{, .}\DecValTok{95}\NormalTok{),}
    \FunctionTok{c}\NormalTok{(.}\DecValTok{01}\NormalTok{, .}\DecValTok{5}\NormalTok{, .}\DecValTok{99}\NormalTok{)}
\NormalTok{  )}
\NormalTok{)}
\end{Highlighting}
\end{Shaded}

with \texttt{records} being the fully annotated data frame or a file
path. \texttt{prev\_classification} can be used to import a different
classification file with prelabeled records.

The hyperparameters are the following:

\begin{itemize}
\tightlist
\item
  resample: whether to use bootstrapping to increase variability between
  the ensemble models;
\item
  n\_init: number of initially labelled records to train the machine
  learning model on;
\item
  n\_models: the number of models to fit and then average to decrease
  the variance due to random sampling in the MCMC algorithm. It reduces
  the likelihood of extreme predictions;
\item
  pos\_mult: the oversampling rate of positive records. Useful to
  improve sensitivity;
\item
  pred\_quants: the distribution quantiles of each PPD, used to define
  the uncertainty zone and identify records to review.
\end{itemize}

Once the grid search is performed, the function
\texttt{analyse\_grid\_search()} can be used to analyse its results.
This function uses a partition tree analysis to identify hyperparameter
sets (\textbf{performance clusters}) with different mean performance
scores. As said before, the default score is Sensitivity x Efficiency;
alternatives are the Positive Rate (i.e., positive records found over
total records) or Positive Rate x Sensitivity.\\
The function produces a data frame with the score and the cluster of
reference of each combination of parameters, the best hyperparameter set
(i.e., the most efficient set among the most sensitive in the best
cluster) and the best set for each cluster. Also, the function creates a
plot to visualise each hyperparameter's impact on performance.

Once chosen, the hyperparameters can be passed to
\texttt{enrich\_annotation\_file()}.\\
Mind that the grid search process may take many days, even on powerful
computers, so we suggest sticking to the default set of parameters when
using \texttt{enrich\_annotation\_file()}.\\

\maketitle

{
\setcounter{tocdepth}{2}
\tableofcontents
}
\newpage

\hypertarget{posterior-predictive-distributions}{%
\section{Posterior predictive
distributions}\label{posterior-predictive-distributions}}

Figure 1 show the mixture of the PPD of the probability of a positive
match, grouped by label (only records manually labelled or reviewed are
considered). The posterior samples for each record were extracted and
joined into a global distribution; on this distribution, density was
computed on the logit scale and then logistic transformed for displaying
on a \([0 - 1]\) scale. The purple ridges show the distribution of the
still unlabelled records.\\
For each iteration, the thresholds which define the ``uncertainty
zone'', i.e, the lower and the upper range of 98\% PrI for the positive
and negative records respectively, are shown. Records whose 98\% PrI
intersects the uncertainty zone requires manual review.\\
Notice how the positive and negative record densities tend to
increasingly overlap at each iteration; meanwhile the distribution of
the records to be reviewed shrinks and shifts towards the negative side,
as positive records get found and labeled.\\
In Session 2, as the number of negative records reviewed increased, also
already positively labelled records re-entered the uncertainty zone;
this is due to the baseline positivity rate decreasing as the number of
negatives in the training data increased, forcing to review dubious
records that may have been mislabelled.

\begin{figure}
\centering
\includegraphics{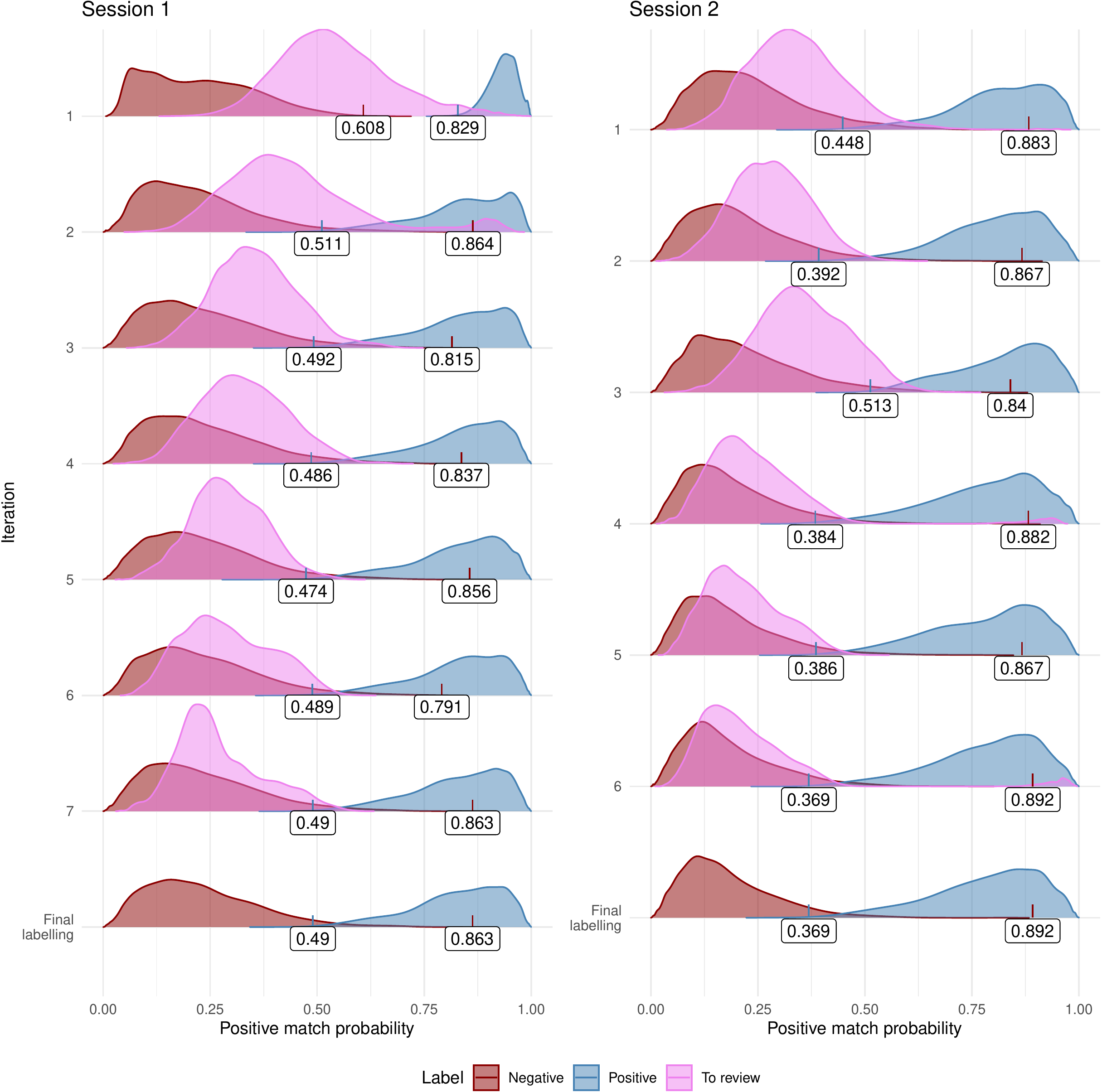}
\caption{\textbf{Figure 1}. Mixture predictive distribution of the
probability of a positive match, grouped by labelling status.}
\end{figure}

\hypertarget{list-of-terms-relevant-for-prediction}{%
\section{List of terms relevant for
prediction}\label{list-of-terms-relevant-for-prediction}}

In table 1 and 2 are listed the 50 more relevant terms used by the BART
algorithm to discriminate between positive and negative records, for
Session 1 and 2 (see Methods). Term importance (``Inclusion rate'' in
the tables) is defined as the ensemble average inclusion rate in
posterior trees over 10,000 total term inclusions, while the Inclusion
Stability (IS) is the ratio of the average inclusion rate over its
standard deviation among the ensemble models. The symbol ``\textbar{}''
in the terms indicate redundant terms while ``\&'' indicate nc-grams.
The component in which the term was used is reported in the leftmost
column.\\
For each term, we added its linear association with a positive label
estimated through a Poisson regression, reporting it as Relative Risk
(RR) and its statistical significance (Statistic) measured as number of
standard errors, s.e., of the terms. A strong BART score with a
Statistic close to zero identify terms whose effect is highly non-linear
(e.g., they are relevant only in the presence of other terms).

\begin{table}[!h]

\caption{\label{tab:var imp}\textbf{Table 1}. Term importance at the end of Session 1.}
\centering
\resizebox{\linewidth}{!}{
\begin{tabular}[t]{llllll}
\toprule
Component & Term & Inclusion rate & IS & RR & Statistic\\
\midrule
\cellcolor{gray!6}{Keyword} & \cellcolor{gray!6}{Patient Transport} & \cellcolor{gray!6}{61.2} & \cellcolor{gray!6}{3.77} & \cellcolor{gray!6}{99.1} & \cellcolor{gray!6}{21.3}\\
Abstract & Transfer & 57 & 3.93 & 22.5 & 15.4\\
\cellcolor{gray!6}{Title} & \cellcolor{gray!6}{Network} & \cellcolor{gray!6}{56.5} & \cellcolor{gray!6}{2.91} & \cellcolor{gray!6}{18} & \cellcolor{gray!6}{14.2}\\
Abstract & Network \& Patient & 54.2 & 4.66 & 26.3 & 15.2\\
\cellcolor{gray!6}{Author} & \cellcolor{gray!6}{Donker T} & \cellcolor{gray!6}{53.5} & \cellcolor{gray!6}{4.56} & \cellcolor{gray!6}{159} & \cellcolor{gray!6}{16.5}\\
\addlinespace
Abstract & Worker & 50 & 3.33 & 0.421 & -1.21\\
\cellcolor{gray!6}{Keyword} & \cellcolor{gray!6}{Hospitals} & \cellcolor{gray!6}{49.8} & \cellcolor{gray!6}{4.31} & \cellcolor{gray!6}{27.8} & \cellcolor{gray!6}{16.5}\\
Abstract & Movement & 47.8 & 2.7 & 27.2 & 15\\
\cellcolor{gray!6}{Title} & \cellcolor{gray!6}{Spread} & \cellcolor{gray!6}{46.6} & \cellcolor{gray!6}{2.25} & \cellcolor{gray!6}{16.2} & \cellcolor{gray!6}{12.1}\\
Abstract & Facility & 45 & 2.22 & 19.6 & 14.8\\
\addlinespace
\cellcolor{gray!6}{Keyword} & \cellcolor{gray!6}{Orange County} & \cellcolor{gray!6}{44.3} & \cellcolor{gray!6}{3.19} & \cellcolor{gray!6}{199} & \cellcolor{gray!6}{17.2}\\
Abstract & Conduct & 42.6 & 3.7 & 0.221 & -2.57\\
\cellcolor{gray!6}{Abstract} & \cellcolor{gray!6}{Patient} & \cellcolor{gray!6}{42} & \cellcolor{gray!6}{3.61} & \cellcolor{gray!6}{27.6} & \cellcolor{gray!6}{7.23}\\
Abstract & Perform & 41.9 & 2.38 & 0.342 & -2.55\\
\cellcolor{gray!6}{Title} & \cellcolor{gray!6}{Hospital} & \cellcolor{gray!6}{39} & \cellcolor{gray!6}{1.95} & \cellcolor{gray!6}{12.5} & \cellcolor{gray!6}{12.5}\\
\addlinespace
Abstract & Regional & 38.9 & 3.08 & 21.7 & 14.9\\
\cellcolor{gray!6}{Abstract} & \cellcolor{gray!6}{Agent} & \cellcolor{gray!6}{38.1} & \cellcolor{gray!6}{2.74} & \cellcolor{gray!6}{4.36} & \cellcolor{gray!6}{6.28}\\
Abstract & California & 37.3 & 2.6 & 38 & 12.6\\
\cellcolor{gray!6}{Title} & \cellcolor{gray!6}{Transfer} & \cellcolor{gray!6}{36.6} & \cellcolor{gray!6}{3.54} & \cellcolor{gray!6}{27} & \cellcolor{gray!6}{11.8}\\
Keyword & Patient Transfer | Patient Transfer \& Patient Transport & 36.6 & 2.16 & 164 & 2\\
\addlinespace
\cellcolor{gray!6}{Abstract} & \cellcolor{gray!6}{Finding} & \cellcolor{gray!6}{33.2} & \cellcolor{gray!6}{1.78} & \cellcolor{gray!6}{0.372} & \cellcolor{gray!6}{-2.35}\\
Title & Outbreak & 32.9 & 2.91 & 3.4 & 3.51\\
\cellcolor{gray!6}{Abstract} & \cellcolor{gray!6}{Collect} & \cellcolor{gray!6}{32.4} & \cellcolor{gray!6}{1.82} & \cellcolor{gray!6}{0.408} & \cellcolor{gray!6}{-1.95}\\
Title & Regional & 32.3 & 2.68 & 44.2 & 14.2\\
\cellcolor{gray!6}{Abstract} & \cellcolor{gray!6}{Network} & \cellcolor{gray!6}{31.6} & \cellcolor{gray!6}{3.09} & \cellcolor{gray!6}{12.8} & \cellcolor{gray!6}{11.7}\\
\addlinespace
Abstract & Resistant & 31.6 & 2.55 & 11 & 11.2\\
\cellcolor{gray!6}{Abstract} & \cellcolor{gray!6}{Outcome} & \cellcolor{gray!6}{31.2} & \cellcolor{gray!6}{1.93} & \cellcolor{gray!6}{0.178} & \cellcolor{gray!6}{-2.95}\\
Abstract & Discharge & 31.1 & 3.79 & 9.99 & 9.02\\
\cellcolor{gray!6}{Abstract} & \cellcolor{gray!6}{2014} & \cellcolor{gray!6}{30.3} & \cellcolor{gray!6}{1.84} & \cellcolor{gray!6}{0.588} & \cellcolor{gray!6}{-1.04}\\
Abstract & Practice & 29.5 & 2.47 & 0.508 & -1.33\\
\addlinespace
\cellcolor{gray!6}{Abstract} & \cellcolor{gray!6}{Culture} & \cellcolor{gray!6}{28.9} & \cellcolor{gray!6}{2.14} & \cellcolor{gray!6}{0.378} & \cellcolor{gray!6}{-1.66}\\
Abstract & Positive & 28.8 & 1.96 & 0.346 & -2.52\\
\cellcolor{gray!6}{Abstract} & \cellcolor{gray!6}{Gene} & \cellcolor{gray!6}{28.3} & \cellcolor{gray!6}{1.29} & \cellcolor{gray!6}{0.00000024} & \cellcolor{gray!6}{-0.0415}\\
Abstract & Disease & 28 & 2.42 & 0.365 & -3.7\\
\cellcolor{gray!6}{Keyword} & \cellcolor{gray!6}{Enterococci} & \cellcolor{gray!6}{27.5} & \cellcolor{gray!6}{1.92} & \cellcolor{gray!6}{25.2} & \cellcolor{gray!6}{6.33}\\
\addlinespace
Abstract & Month & 27.3 & 1.86 & 0.288 & -2.44\\
\cellcolor{gray!6}{Abstract} & \cellcolor{gray!6}{Healthcare \& Facility} & \cellcolor{gray!6}{26.2} & \cellcolor{gray!6}{1.74} & \cellcolor{gray!6}{34.9} & \cellcolor{gray!6}{17}\\
Abstract & Prevalence & 26.2 & 1.71 & 4.08 & 6.69\\
\cellcolor{gray!6}{Abstract} & \cellcolor{gray!6}{Effort} & \cellcolor{gray!6}{26.1} & \cellcolor{gray!6}{1.22} & \cellcolor{gray!6}{6.64} & \cellcolor{gray!6}{8.57}\\
Abstract & Length & 25.3 & 1.44 & 6.1 & 6.45\\
\addlinespace
\cellcolor{gray!6}{Keyword} & \cellcolor{gray!6}{System} & \cellcolor{gray!6}{25} & \cellcolor{gray!6}{1.43} & \cellcolor{gray!6}{11.9} & \cellcolor{gray!6}{3.47}\\
Abstract & Laboratory & 24.4 & 1.38 & 0.442 & -1.39\\
\cellcolor{gray!6}{Keyword} & \cellcolor{gray!6}{Resistant Staphylococcus Aureus} & \cellcolor{gray!6}{24.4} & \cellcolor{gray!6}{1.59} & \cellcolor{gray!6}{22.8} & \cellcolor{gray!6}{11.2}\\
Abstract & Clinical & 23.6 & 2.1 & 0.421 & -2.9\\
\cellcolor{gray!6}{Abstract} & \cellcolor{gray!6}{Dataset} & \cellcolor{gray!6}{22.5} & \cellcolor{gray!6}{1.47} & \cellcolor{gray!6}{8.75} & \cellcolor{gray!6}{6.5}\\
\addlinespace
Abstract & Development & 22.2 & 1.22 & 0.148 & -2.68\\
\cellcolor{gray!6}{Abstract} & \cellcolor{gray!6}{Hand} & \cellcolor{gray!6}{22} & \cellcolor{gray!6}{1.97} & \cellcolor{gray!6}{0.822} & \cellcolor{gray!6}{-0.335}\\
Keyword & Pathogen Transmission & 21.8 & 3.26 & 67.9 & 7.2\\
\cellcolor{gray!6}{Keyword} & \cellcolor{gray!6}{Cross Infection \& Humans \& Transmission} & \cellcolor{gray!6}{21.7} & \cellcolor{gray!6}{1.4} & \cellcolor{gray!6}{31.1} & \cellcolor{gray!6}{15.5}\\
Abstract & Flow & 21.4 & 1.14 & 4.07 & 4.56\\
\bottomrule
\end{tabular}}
\end{table}

\begin{table}[!h]

\caption{\label{tab:var imp}\textbf{Table 2}. Term importance at the end of Session 2.}
\centering
\resizebox{\linewidth}{!}{
\begin{tabular}[t]{llllll}
\toprule
Component & Term & Inclusion rate & IS & RR & Statistic\\
\midrule
\cellcolor{gray!6}{Keyword} & \cellcolor{gray!6}{Patient Transport} & \cellcolor{gray!6}{61.2} & \cellcolor{gray!6}{3.77} & \cellcolor{gray!6}{99.1} & \cellcolor{gray!6}{21.3}\\
Abstract & Transfer & 57 & 3.93 & 22.5 & 15.4\\
\cellcolor{gray!6}{Title} & \cellcolor{gray!6}{Network} & \cellcolor{gray!6}{56.5} & \cellcolor{gray!6}{2.91} & \cellcolor{gray!6}{18} & \cellcolor{gray!6}{14.2}\\
Abstract & Network \& Patient & 54.2 & 4.66 & 26.3 & 15.2\\
\cellcolor{gray!6}{Author} & \cellcolor{gray!6}{Donker T} & \cellcolor{gray!6}{53.5} & \cellcolor{gray!6}{4.56} & \cellcolor{gray!6}{159} & \cellcolor{gray!6}{16.5}\\
\addlinespace
Abstract & Worker & 50 & 3.33 & 0.421 & -1.21\\
\cellcolor{gray!6}{Keyword} & \cellcolor{gray!6}{Hospitals} & \cellcolor{gray!6}{49.8} & \cellcolor{gray!6}{4.31} & \cellcolor{gray!6}{27.8} & \cellcolor{gray!6}{16.5}\\
Abstract & Movement & 47.8 & 2.7 & 27.2 & 15\\
\cellcolor{gray!6}{Title} & \cellcolor{gray!6}{Spread} & \cellcolor{gray!6}{46.6} & \cellcolor{gray!6}{2.25} & \cellcolor{gray!6}{16.2} & \cellcolor{gray!6}{12.1}\\
Abstract & Facility & 45 & 2.22 & 19.6 & 14.8\\
\addlinespace
\cellcolor{gray!6}{Keyword} & \cellcolor{gray!6}{Orange County} & \cellcolor{gray!6}{44.3} & \cellcolor{gray!6}{3.19} & \cellcolor{gray!6}{199} & \cellcolor{gray!6}{17.2}\\
Abstract & Conduct & 42.6 & 3.7 & 0.221 & -2.57\\
\cellcolor{gray!6}{Abstract} & \cellcolor{gray!6}{Patient} & \cellcolor{gray!6}{42} & \cellcolor{gray!6}{3.61} & \cellcolor{gray!6}{27.6} & \cellcolor{gray!6}{7.23}\\
Abstract & Perform & 41.9 & 2.38 & 0.342 & -2.55\\
\cellcolor{gray!6}{Title} & \cellcolor{gray!6}{Hospital} & \cellcolor{gray!6}{39} & \cellcolor{gray!6}{1.95} & \cellcolor{gray!6}{12.5} & \cellcolor{gray!6}{12.5}\\
\addlinespace
Abstract & Regional & 38.9 & 3.08 & 21.7 & 14.9\\
\cellcolor{gray!6}{Abstract} & \cellcolor{gray!6}{Agent} & \cellcolor{gray!6}{38.1} & \cellcolor{gray!6}{2.74} & \cellcolor{gray!6}{4.36} & \cellcolor{gray!6}{6.28}\\
Abstract & California & 37.3 & 2.6 & 38 & 12.6\\
\cellcolor{gray!6}{Title} & \cellcolor{gray!6}{Transfer} & \cellcolor{gray!6}{36.6} & \cellcolor{gray!6}{3.54} & \cellcolor{gray!6}{27} & \cellcolor{gray!6}{11.8}\\
Keyword & Patient Transfer | Patient Transfer \& Patient Transport & 36.6 & 2.16 & 164 & 2\\
\addlinespace
\cellcolor{gray!6}{Abstract} & \cellcolor{gray!6}{Finding} & \cellcolor{gray!6}{33.2} & \cellcolor{gray!6}{1.78} & \cellcolor{gray!6}{0.372} & \cellcolor{gray!6}{-2.35}\\
Title & Outbreak & 32.9 & 2.91 & 3.4 & 3.51\\
\cellcolor{gray!6}{Abstract} & \cellcolor{gray!6}{Collect} & \cellcolor{gray!6}{32.4} & \cellcolor{gray!6}{1.82} & \cellcolor{gray!6}{0.408} & \cellcolor{gray!6}{-1.95}\\
Title & Regional & 32.3 & 2.68 & 44.2 & 14.2\\
\cellcolor{gray!6}{Abstract} & \cellcolor{gray!6}{Network} & \cellcolor{gray!6}{31.6} & \cellcolor{gray!6}{3.09} & \cellcolor{gray!6}{12.8} & \cellcolor{gray!6}{11.7}\\
\addlinespace
Abstract & Resistant & 31.6 & 2.55 & 11 & 11.2\\
\cellcolor{gray!6}{Abstract} & \cellcolor{gray!6}{Outcome} & \cellcolor{gray!6}{31.2} & \cellcolor{gray!6}{1.93} & \cellcolor{gray!6}{0.178} & \cellcolor{gray!6}{-2.95}\\
Abstract & Discharge & 31.1 & 3.79 & 9.99 & 9.02\\
\cellcolor{gray!6}{Abstract} & \cellcolor{gray!6}{2014} & \cellcolor{gray!6}{30.3} & \cellcolor{gray!6}{1.84} & \cellcolor{gray!6}{0.588} & \cellcolor{gray!6}{-1.04}\\
Abstract & Practice & 29.5 & 2.47 & 0.508 & -1.33\\
\addlinespace
\cellcolor{gray!6}{Abstract} & \cellcolor{gray!6}{Culture} & \cellcolor{gray!6}{28.9} & \cellcolor{gray!6}{2.14} & \cellcolor{gray!6}{0.378} & \cellcolor{gray!6}{-1.66}\\
Abstract & Positive & 28.8 & 1.96 & 0.346 & -2.52\\
\cellcolor{gray!6}{Abstract} & \cellcolor{gray!6}{Gene} & \cellcolor{gray!6}{28.3} & \cellcolor{gray!6}{1.29} & \cellcolor{gray!6}{0.00000024} & \cellcolor{gray!6}{-0.0415}\\
Abstract & Disease & 28 & 2.42 & 0.365 & -3.7\\
\cellcolor{gray!6}{Keyword} & \cellcolor{gray!6}{Enterococci} & \cellcolor{gray!6}{27.5} & \cellcolor{gray!6}{1.92} & \cellcolor{gray!6}{25.2} & \cellcolor{gray!6}{6.33}\\
\addlinespace
Abstract & Month & 27.3 & 1.86 & 0.288 & -2.44\\
\cellcolor{gray!6}{Abstract} & \cellcolor{gray!6}{Healthcare \& Facility} & \cellcolor{gray!6}{26.2} & \cellcolor{gray!6}{1.74} & \cellcolor{gray!6}{34.9} & \cellcolor{gray!6}{17}\\
Abstract & Prevalence & 26.2 & 1.71 & 4.08 & 6.69\\
\cellcolor{gray!6}{Abstract} & \cellcolor{gray!6}{Effort} & \cellcolor{gray!6}{26.1} & \cellcolor{gray!6}{1.22} & \cellcolor{gray!6}{6.64} & \cellcolor{gray!6}{8.57}\\
Abstract & Length & 25.3 & 1.44 & 6.1 & 6.45\\
\addlinespace
\cellcolor{gray!6}{Keyword} & \cellcolor{gray!6}{System} & \cellcolor{gray!6}{25} & \cellcolor{gray!6}{1.43} & \cellcolor{gray!6}{11.9} & \cellcolor{gray!6}{3.47}\\
Abstract & Laboratory & 24.4 & 1.38 & 0.442 & -1.39\\
\cellcolor{gray!6}{Keyword} & \cellcolor{gray!6}{Resistant Staphylococcus Aureus} & \cellcolor{gray!6}{24.4} & \cellcolor{gray!6}{1.59} & \cellcolor{gray!6}{22.8} & \cellcolor{gray!6}{11.2}\\
Abstract & Clinical & 23.6 & 2.1 & 0.421 & -2.9\\
\cellcolor{gray!6}{Abstract} & \cellcolor{gray!6}{Dataset} & \cellcolor{gray!6}{22.5} & \cellcolor{gray!6}{1.47} & \cellcolor{gray!6}{8.75} & \cellcolor{gray!6}{6.5}\\
\addlinespace
Abstract & Development & 22.2 & 1.22 & 0.148 & -2.68\\
\cellcolor{gray!6}{Abstract} & \cellcolor{gray!6}{Hand} & \cellcolor{gray!6}{22} & \cellcolor{gray!6}{1.97} & \cellcolor{gray!6}{0.822} & \cellcolor{gray!6}{-0.335}\\
Keyword & Pathogen Transmission & 21.8 & 3.26 & 67.9 & 7.2\\
\cellcolor{gray!6}{Keyword} & \cellcolor{gray!6}{Cross Infection \& Humans \& Transmission} & \cellcolor{gray!6}{21.7} & \cellcolor{gray!6}{1.4} & \cellcolor{gray!6}{31.1} & \cellcolor{gray!6}{15.5}\\
Abstract & Flow & 21.4 & 1.14 & 4.07 & 4.56\\
\bottomrule
\end{tabular}}
\end{table}

\hypertarget{hyperparameters-grid-search}{%
\section{Hyperparameters grid
search}\label{hyperparameters-grid-search}}

Table 3 shows the best hyperparameter set and their performance for each
cluster while figure 2 displays the conditional impact on performance of
each hyperparameter.

The most influential parameters were the number of models in the
ensemble, the oversampling rate and the initial training set size. The
use of bootstrap resampling seemed detrimental.

\begin{figure}
\centering
\includegraphics{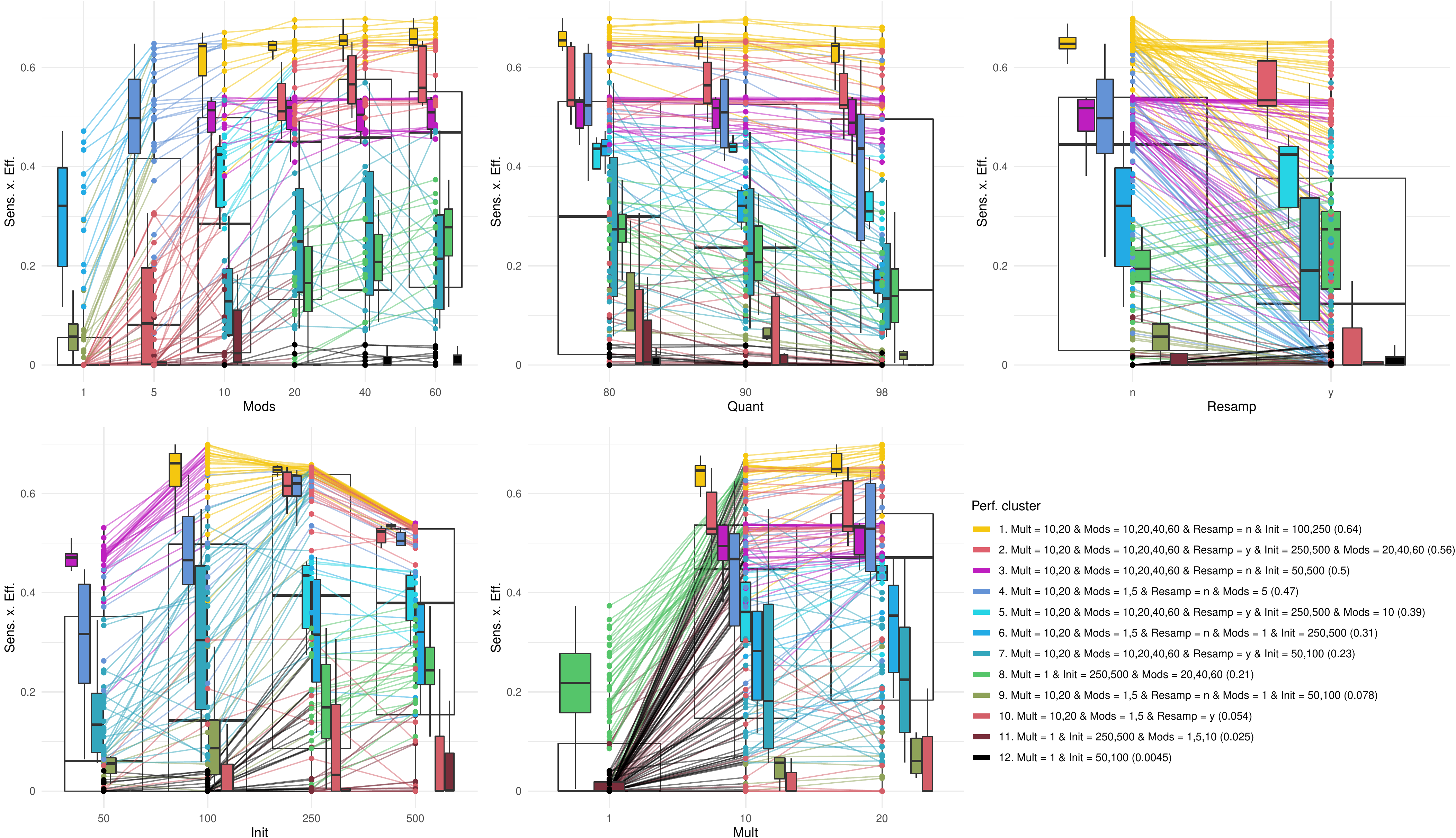}
\caption{\textbf{Figure 2}. Score clusters and impact of single
hyperparameters on engine performance. The performance is measured as
Sensitivity x Efficiency. Each cluster is color coded. Mods: number of
models in the ensemble; Quant: PrI quantiles for the uncertainty zone;
Resamp: bootstrap resampling between models; Init: size of initial
training set; Multi: oversampling rate of positive records}
\end{figure}

\begin{table}[!h]

\caption{\label{tab:hyperparameters}\textbf{Table 3}. Hyperparameter clusters and best cluster subsets. For each cluster, the defining rules and mean Sens. x Eff. is shown, followed by the per-cluster best set results.}
\centering
\resizebox{\linewidth}{!}{
\begin{tabular}[t]{llllllllllll}
\toprule
Cluster (mean score) & Num. iterations & Positive matches & Reviewed records & Sensitivity & Efficiency & Score (Sens. x Eff.) & Num. ensemble models & Uncertainty interval & Resampling & Num. initial training records & Positives oversampling multiplier\\
\midrule
\cellcolor{gray!6}{1. Mult = 10,20 \& Mods = 10,20,40,60 \& Resamp = n \& Init = 100,250 (0.64)} & \cellcolor{gray!6}{7} & \cellcolor{gray!6}{81 / 82} & \cellcolor{gray!6}{462 / 1200} & \cellcolor{gray!6}{98.8\%} & \cellcolor{gray!6}{61.5\%} & \cellcolor{gray!6}{0.608} & \cellcolor{gray!6}{10} & \cellcolor{gray!6}{98} & \cellcolor{gray!6}{n} & \cellcolor{gray!6}{250} & \cellcolor{gray!6}{10}\\
2. Mult = 10,20 \& Mods = 10,20,40,60 \& Resamp = y \& Init = 250,500 \& Mods = 20,40,60 (0.56) & 5 & 82 / 82 & 618 / 1200 & 100\% & 48.5\% & 0.485 & 20 & 80 & y & 250 & 10\\
\cellcolor{gray!6}{3. Mult = 10,20 \& Mods = 10,20,40,60 \& Resamp = n \& Init = 50,500 (0.5)} & \cellcolor{gray!6}{6} & \cellcolor{gray!6}{81 / 82} & \cellcolor{gray!6}{589 / 1200} & \cellcolor{gray!6}{98.8\%} & \cellcolor{gray!6}{50.9\%} & \cellcolor{gray!6}{0.503} & \cellcolor{gray!6}{10} & \cellcolor{gray!6}{98} & \cellcolor{gray!6}{n} & \cellcolor{gray!6}{500} & \cellcolor{gray!6}{10}\\
4. Mult = 10,20 \& Mods = 1,5 \& Resamp = n \& Mods = 5 (0.47) & 8 & 82 / 82 & 1123 / 1200 & 100\% & 6.42\% & 0.0642 & 5 & 98 & n & 50 & 10\\
\cellcolor{gray!6}{5. Mult = 10,20 \& Mods = 10,20,40,60 \& Resamp = y \& Init = 250,500 \& Mods = 10 (0.39)} & \cellcolor{gray!6}{5} & \cellcolor{gray!6}{82 / 82} & \cellcolor{gray!6}{685 / 1200} & \cellcolor{gray!6}{100\%} & \cellcolor{gray!6}{42.9\%} & \cellcolor{gray!6}{0.429} & \cellcolor{gray!6}{10} & \cellcolor{gray!6}{80} & \cellcolor{gray!6}{y} & \cellcolor{gray!6}{250} & \cellcolor{gray!6}{10}\\
\addlinespace
6. Mult = 10,20 \& Mods = 1,5 \& Resamp = n \& Mods = 1 \& Init = 250,500 (0.31) & 5 & 82 / 82 & 874 / 1200 & 100\% & 27.2\% & 0.272 & 1 & 90 & n & 250 & 10\\
\cellcolor{gray!6}{7. Mult = 10,20 \& Mods = 10,20,40,60 \& Resamp = y \& Init = 50,100 (0.23)} & \cellcolor{gray!6}{7} & \cellcolor{gray!6}{82 / 82} & \cellcolor{gray!6}{948 / 1200} & \cellcolor{gray!6}{100\%} & \cellcolor{gray!6}{21\%} & \cellcolor{gray!6}{0.21} & \cellcolor{gray!6}{10} & \cellcolor{gray!6}{90} & \cellcolor{gray!6}{y} & \cellcolor{gray!6}{50} & \cellcolor{gray!6}{20}\\
8. Mult = 1 \& Init = 250,500 \& Mods = 20,40,60 (0.21) & 5 & 82 / 82 & 806 / 1200 & 100\% & 32.8\% & 0.328 & 60 & 80 & y & 250 & 1\\
\cellcolor{gray!6}{9. Mult = 10,20 \& Mods = 1,5 \& Resamp = n \& Mods = 1 \& Init = 50,100 (0.078)} & \cellcolor{gray!6}{6} & \cellcolor{gray!6}{82 / 82} & \cellcolor{gray!6}{1139 / 1200} & \cellcolor{gray!6}{100\%} & \cellcolor{gray!6}{5.08\%} & \cellcolor{gray!6}{0.0508} & \cellcolor{gray!6}{1} & \cellcolor{gray!6}{90} & \cellcolor{gray!6}{n} & \cellcolor{gray!6}{50} & \cellcolor{gray!6}{20}\\
10. Mult = 10,20 \& Mods = 1,5 \& Resamp = y (0.054) & 5 & 82 / 82 & 832 / 1200 & 100\% & 30.7\% & 0.307 & 5 & 80 & y & 250 & 20\\
\addlinespace
\cellcolor{gray!6}{11. Mult = 1 \& Init = 250,500 \& Mods = 1,5,10 (0.025)} & \cellcolor{gray!6}{5} & \cellcolor{gray!6}{82 / 82} & \cellcolor{gray!6}{981 / 1200} & \cellcolor{gray!6}{100\%} & \cellcolor{gray!6}{18.2\%} & \cellcolor{gray!6}{0.182} & \cellcolor{gray!6}{10} & \cellcolor{gray!6}{90} & \cellcolor{gray!6}{y} & \cellcolor{gray!6}{500} & \cellcolor{gray!6}{1}\\
12. Mult = 1 \& Init = 50,100 (0.0045) & 6 & 82 / 82 & 1151 / 1200 & 100\% & 4.08\% & 0.0408 & 20 & 80 & y & 50 & 1\\
\bottomrule
\end{tabular}}
\end{table}

\maketitle

\hypertarget{wos}{%
\subsection{WOS}\label{wos}}

\begin{itemize}
\tightlist
\item
  activate your institution proxy or log in in some way (it is only
  needed to get the API key, then you can log off the proxy)
\item
  go to \url{https://apps.webofknowledge.com/}
\item
  extract the API key from the URL, is denominated by SID
\item
  put the key in the secrets.R file and source it or source Setup.R
\item
  perform the search using the API
\end{itemize}

\hypertarget{ieee}{%
\subsection{IEEE}\label{ieee}}

\begin{itemize}
\tightlist
\item
  get your API key and put it into the secrets.R
\item
  source the file source Setup.R
\item
  perform the search using the API
\end{itemize}

\hypertarget{scopus}{%
\subsection{Scopus}\label{scopus}}

\begin{itemize}
\tightlist
\item
  change the NOT in the query to AND NOT
\item
  go to
  \url{https://www.scopus.com/search/form.uri?display=basic\#basic}
\item
  log in with your institution credential
\item
  set up the year range
\item
  search within: title, abstract, keywords, authors
\item
  select years in order to have just less than 2000 records
\item
  download csv (Excel). Choose ``Citation information'' and ``Abstract
  \& keywords''
\item
  call the files Scopus\#.csv with \# being a sequential number
\item
  perform\_search\_session() will find and parse the files
\end{itemize}

\hypertarget{pubmed}{%
\subsection{Pubmed}\label{pubmed}}

\begin{itemize}
\tightlist
\item
  put your API key in the secrets.R file and source it or source Setup.R
\item
  perform the search using the API
\item
  put the same query into pubmed and download records in the Pubmed
  format.
\item
  only 10k records can be downloaded at once, so use the year filter to
  partition the records
\item
  call the file Pubmed\#.nbib with \# being a sequential number
\item
  perform\_search\_session() will find and parse the files
\end{itemize}

\hypertarget{embase}{%
\subsection{Embase}\label{embase}}

\begin{itemize}
\tightlist
\item
  Login from within the institution or activate remote access
\item
  go to \url{https://www.embase.com/\#advancedSearch/default}
\item
  set your date range
\item
  in the mapping field select all apart from ``Limit to terms indexed in
  article as `major focus'\,''
\item
  in ``Sources'' select Embase
\item
  select records in blocks of 10K
\item
  chose format: Csv - fields by column, output: Full record (may take a
  while to prepare the download and you can download only one batch at
  time)
\item
  call the files Embase\#.csv with \# being a sequential number
\end{itemize}